\documentclass[twocolumn,prl,aps,superscriptaddress]{revtex4}
\usepackage{graphicx}
\usepackage{float}
\usepackage{dcolumn}
\usepackage{bm}
\usepackage{color}
\usepackage{amsmath}
\usepackage{amssymb}
\usepackage{amsfonts}
\usepackage{esint}
\usepackage{times}
\usepackage{xcolor}
\usepackage{phaistos}
\usepackage{braket}
\usepackage{comment}

\usepackage{pdfpages}

\usepackage[colorlinks,linkcolor=blue,anchorcolor=blue,citecolor=blue]{hyperref}

\begin{document}

\title{High-contrast ZZ interaction using superconducting qubits with opposite-sign anharmonicity}

\author{Peng Zhao}\email{shangniguo@sina.com}
\affiliation{National Laboratory of Solid State Microstructures, School of Physics,
Nanjing University, Nanjing 230039, China}

\author{Peng Xu}
\affiliation{National Laboratory of Solid State Microstructures, School of Physics,
Nanjing University, Nanjing 230039, China} \affiliation{Institute of Quantum Information and Technology, Nanjing University of Posts and Telecommunications, Nanjing, 210003, China}\affiliation{State Key Laboratory of Quantum Optics and Devices, Shanxi University, Taiyuan, 030006, China}

\author{Dong Lan}
\affiliation{National Laboratory of Solid State Microstructures, School of Physics,
Nanjing University, Nanjing 230039, China}

\author{Ji Chu}
\affiliation{National Laboratory of Solid State Microstructures, School of Physics,
Nanjing University, Nanjing 230039, China}

\author{Xinsheng Tan}\email{meisen0103@163.com}
\affiliation{National Laboratory of Solid State Microstructures, School of Physics,
Nanjing University, Nanjing 230039, China}

\author{Haifeng Yu}
\affiliation{National Laboratory of Solid State Microstructures, School of Physics,
Nanjing University, Nanjing 230039, China}

\author{Yang Yu}
\affiliation{National Laboratory of Solid State Microstructures, School of Physics,
Nanjing University, Nanjing 230039, China}

\date{\today}

\begin{abstract}
For building a scalable quantum processor with superconducting qubits,
ZZ interaction is of great concern because its residual has a crucial impact to two-qubit gate fidelity.
Two-qubit gates with fidelity meeting the criterion of fault-tolerant quantum computationhave been demonstrated using ZZ
interaction. However, as the performance of quantum processors
improves, the residual static-ZZ can become a performance-limiting factor
for quantum gate operation and quantum error correction. Here, we
introduce a superconducting architecture using qubits with opposite-sign
anharmonicity, a transmon qubit and a C-shunt flux qubit, to address this
issue. We theoretically demonstrate that by coupling the two types of qubits, the
high-contrast ZZ interaction can be realized. Thus, we can
control the interaction with a high on/off ratio to implement two-qubit
CZ gates, or suppress it during two-qubit gate operation using XY
interaction (e.g., an iSWAP gate). The proposed architecture can also be scaled
up to multi-qubit cases. In a fixed coupled system, ZZ crosstalk
related to neighboring spectator qubits could also be heavily suppressed.
\end{abstract}

\maketitle



Engineering a physical system for fault-tolerant quantum computing
demands quantum gates with error rates below the fault-tolerant
threshold, which has been demonstrated in small-sized superconducting
quantum processors \cite{R1}. At present, high-performance
quantum processors with dozens of superconducting qubits have
become available \cite{R2}, but realizing fault-tolerant quantum computing
is still out of reach, mainly because of the heavy overhead needed for
error-correction with state-of-the-art gate performance. Therefore, further improving
gate performance is essential for realizing fault-tolerant
quantum computing with supercondcuting qubits.

With today's superconducting quantum processors, apart from increasing qubit
coherence times, speeding up gates can also fundamentally improve
gate performance. However, there is a speed-fidelity trade-off imposed by parasitic
interactions. Since the current two-qubit gates typically have lower gate
speeds and worse fidelity than single-qubit gates \cite{R3}, this issue is particularly
relevant to two-qubit gates. For implementing a fast two-qubit gates with strong two-qubit
coupling, one of the major parasitic interactions is ZZ coupling, which is mostly
caused by the coupling between higher energy levels of qubits \cite{R4,R5}.
Thus, for qubits with weak anharmonicity, such as transmon qubits \cite{R6} and C-shunt
flux qubits (in single well regions) \cite{R7,R8,R9}, the non-zero parasitic ZZ coupling exists
inherently due to the intrinsic energy level diagrams of qubits. This ZZ interaction has been shown to act
as a double-edged sword for quantum computing: it can be used to implement
high-speed and high-fidelity controlled-Z (CZ) gates \cite{R1,R10,R11,R12}, yet it
can also degrade performance of two-qubit gates through XY interaction \cite{R2,R8,R11,R12,R13,R14,R15,R16,R17}.
Moreover, in fixed coupled multi-qubit systems, such as the one shown
in Fig.1(a), gate operations in the two qubits enclosed by the rectangle involve
six neighboring spectator qubits, and the ZZ coupling related to these
qubits cannot be fully turned off by tuning qubits out of resonance \cite{R1}.
The residual is typically manifested as crosstalk, which results in addressing
errors and phase errors during gate operations and error correction \cite{R18,R19,R20,R21,R22,R23,R24}.
Furthermore, these errors are correlated multi-qubit errors, which are particularly
harmful for realizing a fault-tolerant scheme \cite{R25}. Given the fidelity and performance
limitations related to parasitic ZZ interaction, it is highly
desirable to achieve high-contrast control over this parasitic coupling.

To address this challenge, in this work, we introduce a superconducting architecture
using qubits with opposite-sign anharmonicity. We theoretically demonstrate
our protocol with coupled transmon and C-shunt
flux qubits, which have negative and positive anharmonicity, respectively.
We show that high-contrast ZZ interaction can be achieved by engineering the
system parameters. By utilizing ZZ interaction with a high on/off ratio, we can
implement the CZ gate with a speed higher than that of the traditional setup using
only one type of qubit (e.g., full transmon systems). Parasitic ZZ coupling can also
be deliberately suppressed during two-qubit gate operations using XY
interaction (e.g., an iSWAP gate), while leaving the XY interaction completely intact.
The proposed architecture can be scaled up to multi-qubit cases, and in fixed
coupled systems, ZZ crosstalk related to spectator qubits could also be
heavily suppressed.

\begin{figure}[tbp]
\begin{center}
\includegraphics[width=8cm,height=4.5cm]{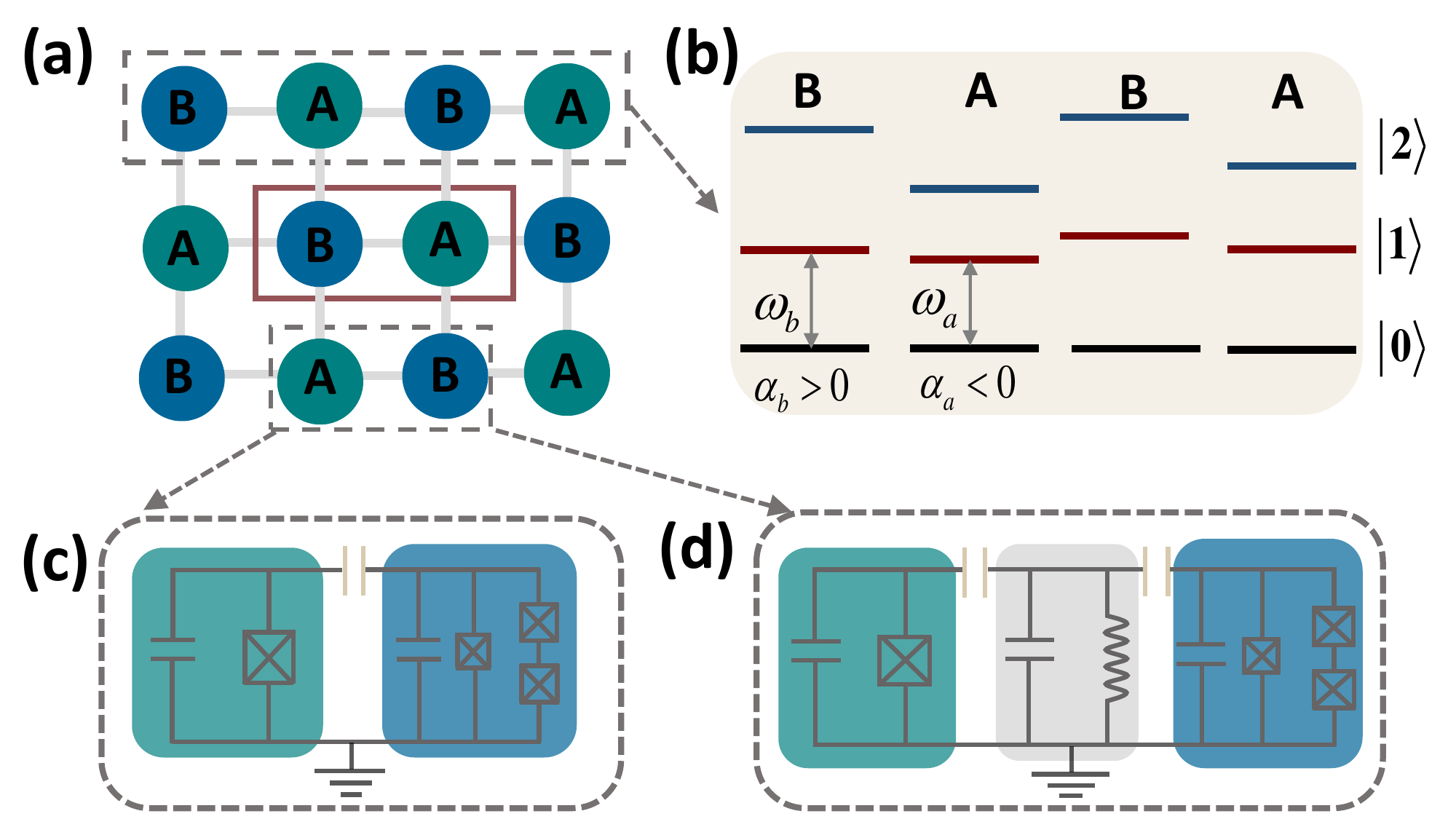}
\end{center}
\caption{(a) Layout of a two-dimensional nearest-neighbor lattice, where circles
at the vertices denote qubits, and gray lines indicate
couplers between adjacent qubits. The lattice consists of two-type
qubits arranged in an -A-B-A-B- pattern in each row and column. The circles
with A and B are qubits with opposite-sign
anharmonicities, and each one can be treated as a three-level anharmonic oscillators (b).
Typically, transmon qubits and C-shunt flux qubits can
be modeled as anharmonic oscillators with negative and positive anharmonicity,
respectively. Qubits can be coupled to each other (c) directly via a
capacitor or (d) indirectly using a resonator.}
\end{figure}


To start, let us consider a superconducting architecture (hereinafter, the AB-type) where two qubits with
opposite-sign anharmonicities are coupled together. The architecture can be treated
as a module that can be easily scaled up to multi-qubit case. In Fig.$\,$1(a),
we show a case of a nearest-neighbor-coupled qubit lattice, where
circles with A and B are two-type qubits with opposite-sign
anharmonicities arranged in an -A-B-A-B- pattern. As shown in Fig.$\,$1(b), both qubits can be
modeled as a three-level (i.e., $|0\rangle,\, |1\rangle,\, |2\rangle$)
anharmonic oscillator for which the Hamiltonian is given as ($\hbar=1$)
\begin{eqnarray}
\begin{aligned}
H_{l}=\omega_{l}q_{l}^{\dagger}q_{l}+\frac{\alpha_{l}}{2}q_{l}^{\dagger}q_{l}(q_{l}^{\dagger}q_{l}-1),
\end{aligned}
\end{eqnarray}
where the subscript $l={a,b}$ labels an anharmonic oscillator with anharmonicity
$\alpha_{l}$ and frequency $\omega_{l}$, and $q_{l}\,(q_{l}^{\dagger})$
is the associated annihilation (creation) operator truncated to the
lowest three-level. Usually, qubits can be coupled directly via a capacitor
or indirectly through a bus resonator, as shown in Fig.$\,$1(c) and 1(d). For clarity
and without loss of generality, unless explicitly mentioned, we focus on the direct-coupled case
in the following discussion, and the
dynamics of two-coupled qubits can be described by the
Hamiltonian $H=H_{a}+H_{b}+H_{I}$, where $H_{I}=g(q_{a}^{\dagger}q_{b}+H.c.)$
describes the inter-qubit coupling with strength $g$.

Before describing our main idea of engineering high-contrast ZZ interaction
in our architecture, we need to first examine the origin of
parasitic ZZ interaction in the traditional setup (hereinafter, the AA-type),
where two transmon qubits are coupled directly. By ignoring higher energy
levels, we can model the transmon qubit as an anharmonic oscillator with
negative anharmonicity \cite{R6}, thus the system Hamiltonian can be expressed
as $H$ with $\alpha_{a,b}<0$. Fig.$\,$2(a) shows numerically calculated energy
levels of the system for anharmonicities $\alpha_{a,b}=-\alpha$
($\alpha/2\pi=250\,\rm MHz$, which is a positive number throughout this work) \cite{R26}.
One can find that there are four avoided crossings, one corresponds to the XY
interaction in the one-excitation manifold (i.e., interaction
$|01\rangle\leftrightarrow|10\rangle$), and the other three (from left to right)
associate with interactions among the two-excitation manifold $\{|11\rangle, |02\rangle, |20\rangle\}$ (i.e., interactions $|11\rangle\leftrightarrow|02\rangle$, $|20\rangle\leftrightarrow|02\rangle$, and $|11\rangle\leftrightarrow|20\rangle$).
The interactions between qubit states $|11\rangle$ and non-qubit states ($|02\rangle,|20\rangle$)
change the energy of state $|\widetilde{11}\rangle$, where $|\widetilde{ij}\rangle$ denotes
the eigenstate of the Hamiltonian $H$ that has the maximum overlap with the bare
state $|ij\rangle$, and the corresponding eigenenergy is $E_{\widetilde{ij}}$, thus
leading to ZZ coupling with strength
\begin{eqnarray}
\begin{aligned}
\zeta=(E_{\widetilde{11}}-E_{\widetilde{01}})-(E_{\widetilde{10}}-E_{\widetilde{00}})=J(\tan\frac{\theta_{b}}{2}-\tan\frac{\theta_{a}}{2}),
\end{aligned}
\end{eqnarray}
where $\tan\theta_{a,b}=2J/(\Delta\pm\alpha_{a,b})$, $\Delta=\omega_{a}-\omega_{b}$ denotes
qubit detuning, and $J=\sqrt{2}g$ is the
coupling strength of $|11\rangle\leftrightarrow|02\rangle\,(|20\rangle)$.
When $J\ll|\Delta\pm\alpha_{a,b}|$, Eq.$\,$(2) can be approximated by $\zeta=J^{2}/(\Delta-\alpha_{b})
-J^{2}/(\Delta+\alpha_{a})$ \cite{R1}.

With the expression in Eq.$\,$(2), there are two terms contributing to ZZ interaction, and each
is independently associated with the interaction $|11\rangle\leftrightarrow|02\rangle\,(|20\rangle)$.
Replacing one of the two
transmon qubits with a qubit for which the value of anharmonicity is comparable but the sign is positive
can destructively interfere the two terms in Eq.$\,$(2), thus heavily suppressing ZZ coupling.
A promising qubit to implementing such a AB-type setup is the C-shunt flux qubit in
single well regime \cite{R7,R8,R9}, where the qubit can be modeled as an
anharmonic oscillator that has positive anharmonicity with magnitude comparable to
that of the transmon qubit \cite{R8,R26}. In Fig.$\,$2(b), we show numerically
calculated energy levels for this AB-type setup with $\alpha_{b}/2\pi=250\rm\,MHz$ \cite{R8}
and keep all other parameters the same as in Fig.$\,$2(a). Compared with the
AA-type setup, the avoided crossing associated with interaction $|01\rangle\leftrightarrow|10\rangle$
is completely intact, but the interaction among two-excitation manifold forms an avoided crossing
with triplets, as shown in the inset of Fig.$\,$2(b). At the triple degeneracy
point, the eigenstates are $(|02\rangle+|20\rangle-\sqrt{2}|11\rangle)/2$,
$(|02\rangle-|20\rangle)/\sqrt{2}$, $(|02\rangle+|20\rangle+\sqrt{2}|11\rangle)/2$,
with the corresponding energies of $E_{11}-\sqrt{2}J$, $E_{11}$,
and $E_{11}+\sqrt{2}J$ \cite{R31}.

\begin{figure}[t]
\begin{center}
\includegraphics[width=8cm, height=6cm]{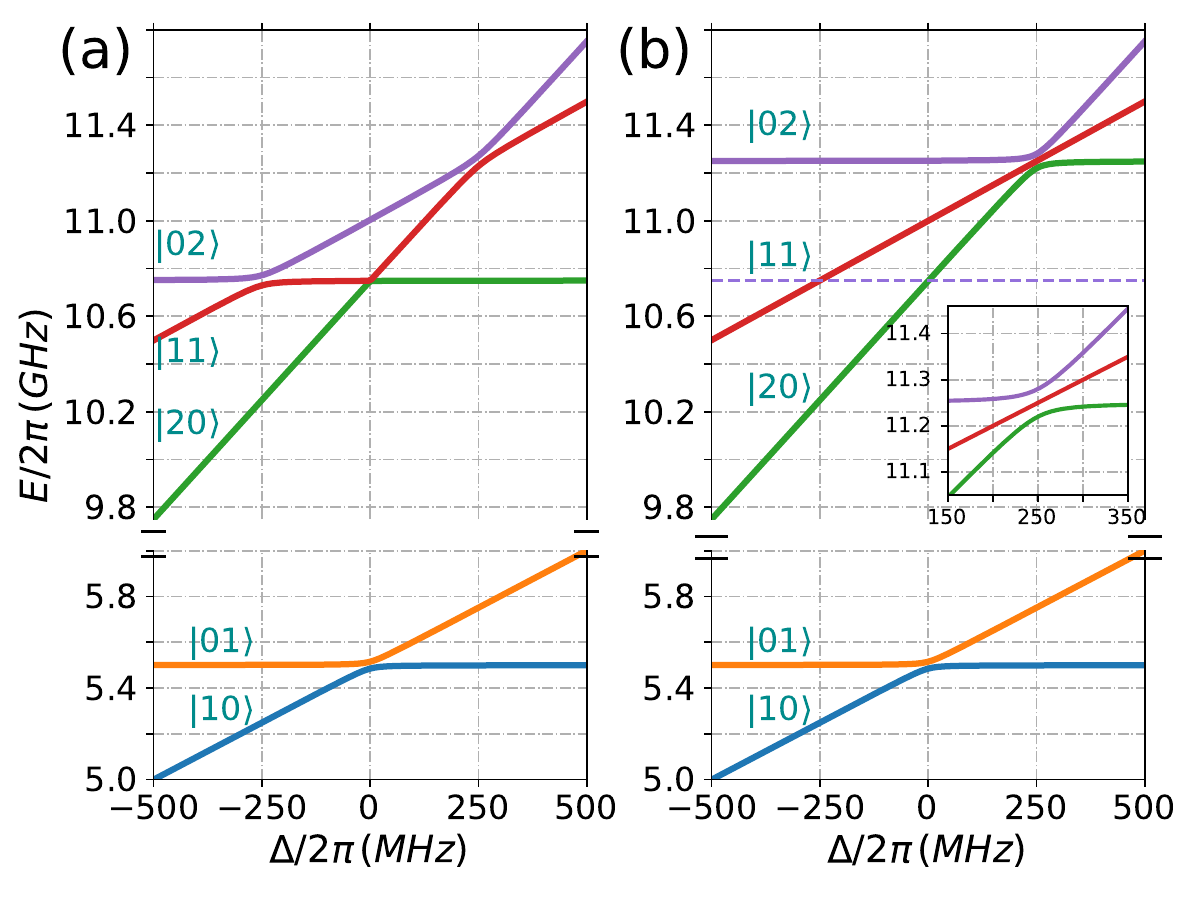}
\end{center}
\caption{ Numerical calculation of the energy levels of the
coupled qubit system as a function of the qubit detuning $\Delta=\omega_{a}-\omega_{b}$
for qubit frequency $\omega_{b}/(2\pi)=5.5\,\rm GHz$, anharmonicity
$\alpha/2\pi=250\,\rm MHz$ \cite{R8,R26}, and coupling strength $g/2\pi=15\,\rm MHz$.
(a) Qubits with same-sign anharmonicity $\alpha_{a(b)}=-\alpha$; (b) Qubits
with opposite-sign anharmonicity $\alpha_{a(b)}=\mp\alpha$.
The inset highlights the triple degeneracy point in the two-excitation
manifold $\{|02\rangle$,$|20\rangle$,$|11\rangle\}$.}
\end{figure}


Fig.$\,$3(a) shows the numerical result of ZZ coupling strength as a function of
qubit detuning $\Delta$ in the AB-type setup. The result for the AA-type setup
is also shown for comparison. In the AB-type setup, ZZ coupling is completely removed
away from the triple degeneracy point at $\Delta=\alpha_{b}$, while for regions close to the
degeneracy point, coupling is preserved, and the strength at the degeneracy
point is larger than that of the AA-type setup ($2g$ vs $\sqrt{2}g$) \cite{R31}. In
Fig.$\,$3(b), we have also shown ZZ coupling strength as a function of the
anharmonicity asymmetry $\delta_{\alpha}=|\alpha_{b}|-|\alpha_{a}|$ for typical
coupling strength $g/2\pi=15\,\rm MHz$ and qubit detuning $\Delta/2\pi=-150\,\rm MHz$.
One can find that the ZZ coupling strength is suppressed below $0.7\,\rm{MHz}$ for
the anharmonicity asymmetry around $-100\sim 600\,\rm{MHz}$. In addition,
since the anharmonicities of both types of qubits (the transmon qubit and
the C-shunt flux qubit) depend primarily on geometric circuit parameters,
the typical anharmonicity asymmetry $\delta_{\alpha}$ around $-20\sim 20\,\rm{MHz}$
could be achieved with current qubit fabrication techniques \cite{R26}. In this
case, ZZ coupling strength could be further suppressed below $60\,\rm KHz$, as
shown in the inset of Fig.$\,$3(b), whereas for the traditional setup, the typical
strength of the residual ZZ coupling is about $5\,\rm{MHz}$, as shown in Fig.$\,$3(a).

For a more comprehensive analysis of ZZ coupling in the AB-type setup, we
explore the full parameter range in Fig.$\,$3(c) with varying qubit detuning $\Delta$
and ahharmonicity asymmetry $\delta_{\alpha}$. We identify three regions
in parameter space with prominent characteristic. The two lighter
regions indicate that ZZ coupling becomes strong when qubit
detuning approaches qubit anharmonicity, i.e., $\Delta=\mp\alpha_{a,b}$, and
the intersection region corresponds to the triple degeneracy point. The darker region
shows where ZZ coupling is heavily suppressed and is zero for $\delta_{\alpha}=0$.
In Fig.$\,$3(d), we show the result for an indirect-coupled case, where
qubits are coupled via a resonator \cite{R32}. In this case, the strength of the effective
inter-qubit coupling depends on qubit detuning, thus the zero ZZ coupling point
depends not only on the anharmonicity asymmetry, but also on the qubit detuning.

\begin{figure}[tbp]
\begin{center}
\includegraphics[width=8cm, height=6cm]{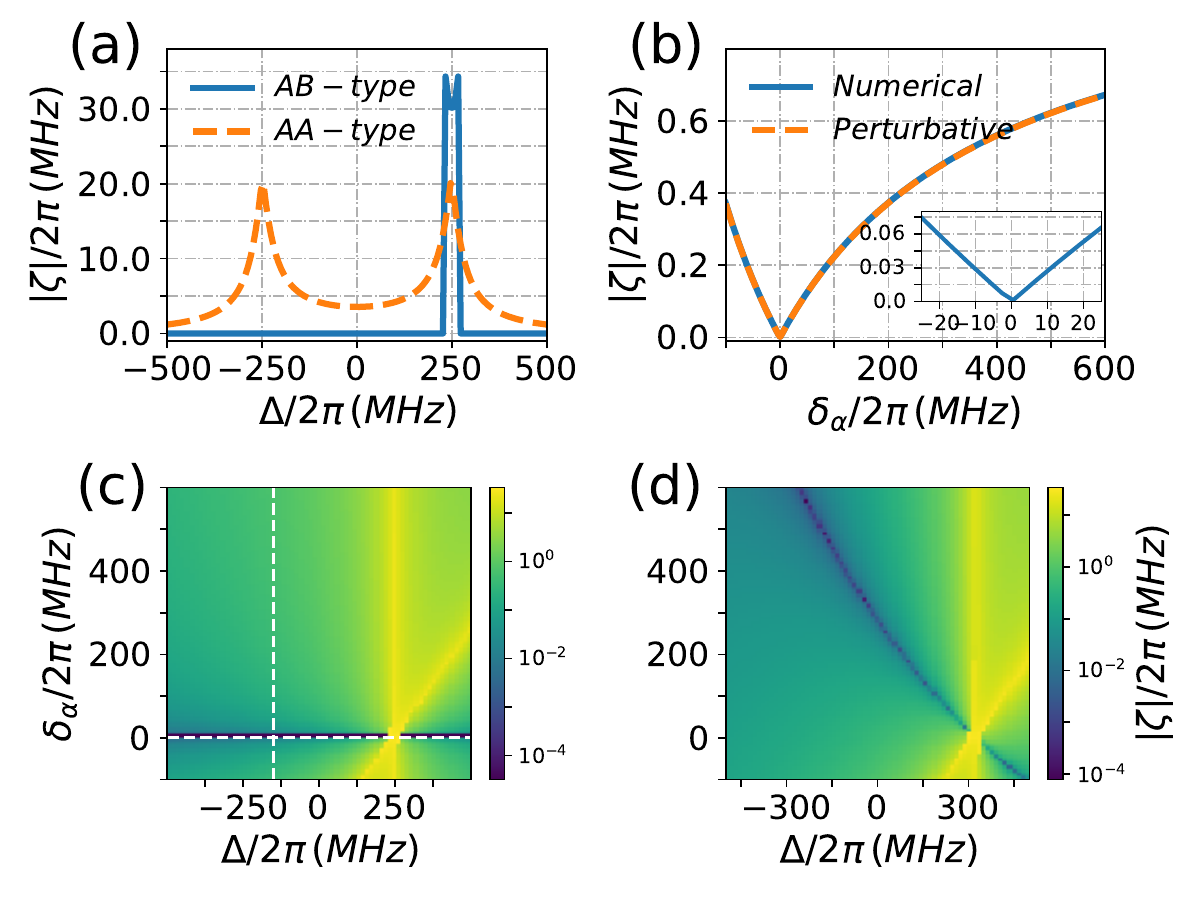}
\end{center}
\caption{ Numerical results for ZZ coupling strength $|\zeta|$.
(a) $|\zeta|$ versus qubit detuning $\Delta$ for anharmonicity
asymmetry $\delta_{\alpha}=|\alpha_{b}|-|\alpha_{a}|=0$, where the dashed line shows
result for the AA-type. (b) $|\zeta|$ versus $\delta_{\alpha}$ for $\Delta/2\pi=-150\,\rm MHz$,
where the dashed line shows perturbational result. The inset shows that $|\zeta|$
could be suppressed below $60\,\rm KHz$ with the typical anharmonicity
asymmetry ($-20\sim 20\,\rm{MHz}$). (c) $|\zeta|$ versus $\Delta$
and $\delta_{\alpha}$. Horizontal (vertical) cut through (c) denotes the result
plotted in a (b). (d) ZZ coupling in a system comprising two
qubits that are coupled via a resonator \cite{R32}.}
\end{figure}


\begin{figure}[tbp]
\begin{center}
\includegraphics[width=8cm,height=5cm]{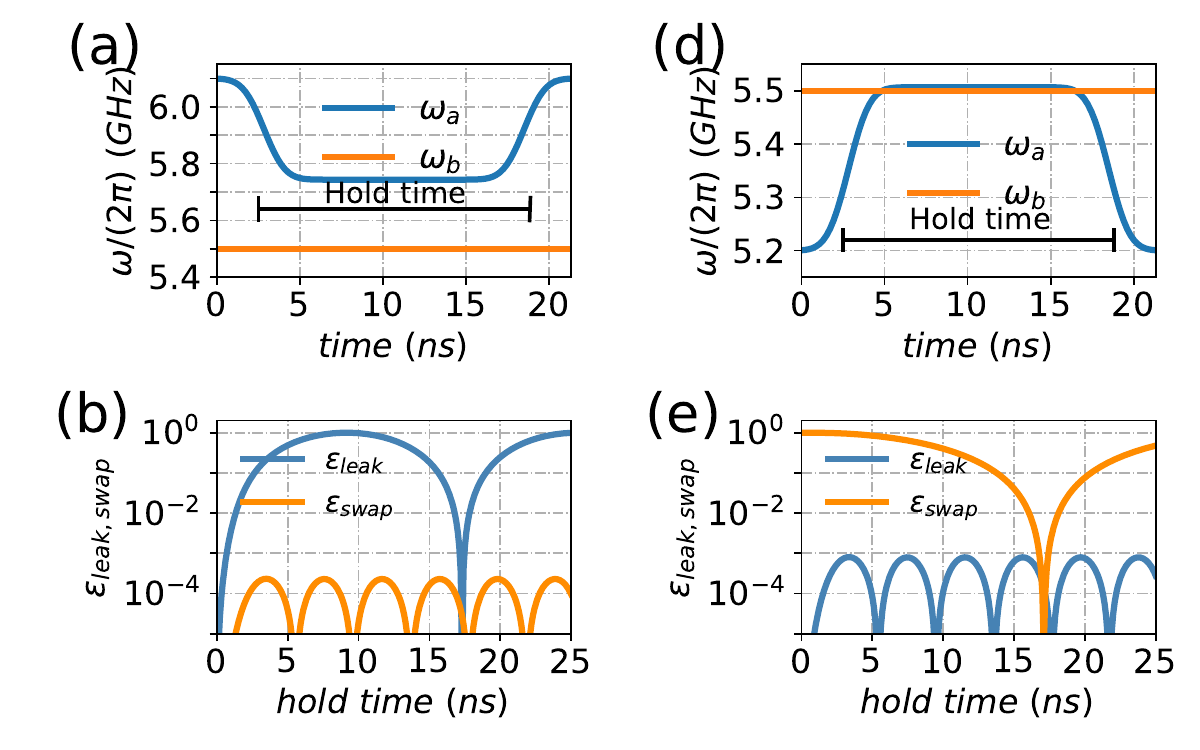}
\includegraphics[width=8cm,height=3.5cm]{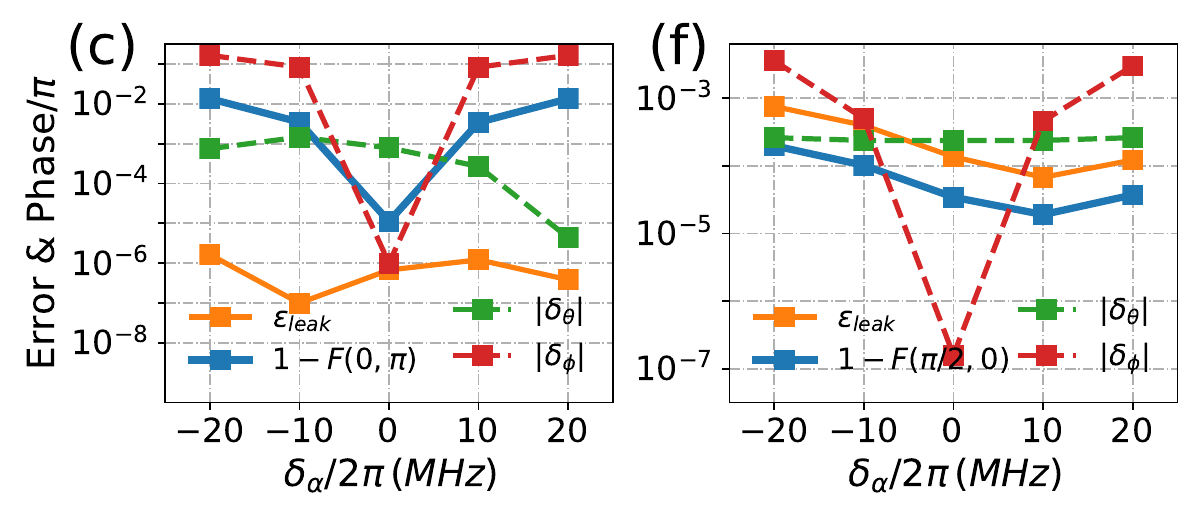}
\end{center}
\caption{Numerical results for CZ gate and iSWAP gate implementation
in our architecture. The system parameters used are same as in Fig.$\,$2(b).
(a),(d) Typical pulses with small overshoots for
realizing CZ gates and iSWAP gates, where the full width at half maximum
is defined as hold time. (b),(e) Leakage $\varepsilon_{\rm laek}$ and swap
error $\varepsilon_{\rm swap}$ versus hold time for system with anharmonicity asymmetry
$\delta_{\alpha}=0$ and optimal overshoots \cite{R35}.
(c),(f) Gate error $1-F$ versus typical anharmonicity asymmetry. The phase
error $\delta_{\theta},\delta_{\phi}$ with respect to the ideal phase
parameters ($0,\pi$ for CZ gate, and $\pi/2,0$ for iSWAP gate) and leakage error
$\varepsilon_{\rm leak}$ are also presented for identifying the major source of error.}
\end{figure}

Having shown high contrast ZZ interaction in the AB-type setup, we now turn to
study the implementation of two-qubit gates with a diabatic scheme in this setup \cite{R4,R11}.
Here, we focus on the direct-coupled system with
always-on interactions described by the Hamiltonian $H$ with $\alpha_{a}<0$ and
$\alpha_{b}>0$, but the method is generalizable to other coupled systems \cite{R32}. For
illustration purposes and easy reference, we use the same parameters as those in Fig.$\,$2(b).
In this case, during the gate operations, the frequency of qubit $b$ remains at
its parking point, while the frequency of qubit $a$ changes from its parking point to the
interaction point and then back, according to a time-dependent function \cite{R35,R38},
as shown in Fig.$\,$4(a) or 4(d) where the full width at half maximum is defined as
hold time. We note that at the parking (idling) point where the inter-qubit coupling is effectively
turned off, the logical basis state $|\overline{ij}\rangle$
is defined as the eigenstates of the system biased at this point \cite{R31,R38},
which is adiabatically connected to the bare state $|ij\rangle$.
Expressed in the logical basis, the target gate operations can be expressed as
\begin{eqnarray}
\begin{aligned}
U(\theta,\phi)=e^{-i(|\overline{01}\rangle\langle \overline{10}|+|\overline{10}\rangle\langle \overline{01}|)
\theta}e^{-i|\overline{11}\rangle\langle \overline{11}|\phi},
\end{aligned}
\end{eqnarray}
where $\theta$ denotes the swap angle associated with the bare exchange
interaction $|01\rangle\leftrightarrow|10\rangle$, and $\phi$ represents
the conditional phase resulting from ZZ coupling. To quantify the intrinsic performance of
the implemented gate operation, we use the metric of state-average
gate fidelity $F(\theta,\phi)=[{\rm Tr}(U^{\dagger}U)+|{\rm Tr}(U(\theta,\phi)^{\dagger}U)|^{2}]/20$ \cite{R39},
where $U$ is the actual evolution operator (ignoring the decoherence process) up
to single qubit phase gates.

We first consider the implementation of the CZ gate $U(0,\pi)$, and the main idea is
as follows. By tuning the frequency of qubit $a$ from its parking point $\omega_{P}=6.1\,\rm GHz$
to the interaction point $\omega_{I}=\omega_{b}+\alpha$ according to the time-dependent
function shown in Fig.$\,$4(a), the CZ gate can be realized after a full Rabi
oscillation between $|11\rangle$ and $[|02\rangle+|20\rangle]/\sqrt{2}$. As
mentioned before, the Rabi rate is larger than in the AA-type setup ($2g$ vs $\sqrt{2}g$),
thus allowing a higher gate speed. We note that an additional small overshoot to the
interaction frequency $\omega_{I}$ is critical to optimize the leakage to non-qubit
states \cite{R11,R35}. By taking the optimal overshoot and initializing the system
in states $|\overline{11}\rangle$ and $|\overline{01}\rangle$, Fig.4(b) shows the leakage
$\varepsilon_{\rm leak}=1-P_{\bar{11}}$ ($P_{\bar{ij}}$ denotes the population in the logical
state $|\overline{ij}\rangle$ at the end of the gate operations) and swap error
$\epsilon_{\rm swap}=1-P_{\bar{01}}$ as a function of the hold time. In present
system with fixed inter-qubit coupling, it is nearly impossible to have an optimal hold
time to simultaneously minimize the swap error and leakage, as shown in Fig.4(b). Thus,
here, we choose to minimize the leakage, and find that with a
hold time of $17.3\,\rm ns$, a CZ gate with fidelity above $99.999\%$
can be achieved, and both the leakage and swap error can be suppressed to below $10^{-4}$.
However, as shown in Fig.$4$(c), when the system is considered with typical anharmonicity
asymmetry $\delta_{\alpha}$, gate fidelity worsens. In order to identify the performance
limiting factors, we extract phase error $\delta_{\theta},\,\delta_{\phi}$ with
respect to the ideal phase parameters $\theta=0, \,\phi=\pi$ for CZ gate and the leakage, and find that
in the current case, gate error primarily result from conditional phase error
$\delta_{\phi}$. This can be explained by the fact that in systems with typical anharmonicity
asymmetry, the resonance condition for having a full Rabi oscillation between $|11\rangle$ and
non-qubit state breaks down. Off-resonance Rabi oscillation is thus presented, causing conditional
phase error \cite{R35}.

An iSWAP gate $U(\pi/2,0)$ can be realized by tuning the two qubits into resonance
according to the control pulse shown in Fig.$\,$4(d). Given an optimal
overshoot with respect to the interaction point $\omega_{a}=\omega_{b}$, an iSWAP gate
with fidelity exceeding $99.99\%$ can be realized with a hold time of $17.1\,\rm ns$ \cite{R40}, and
leakage $\varepsilon_{\rm leak}$ and swap error $\varepsilon_{\rm swap}=P_{\bar{01}}$ can
be suppressed to below $10^{-4}$, as shown in Fig.$\,$4(e). As shown in Fig.$\,$4(f), we also study the
effect of anharmonicity asymmetry on gate fidelity, and find that gate fidelity
in excess of $99.98\%$ can be achieved for system with typical anharmonicity asymmetry.
By extracting phase error $\delta_{\theta},\,\delta_{\phi}$ and leakage for the iSWAP gate,
we find that leakage error becomes the major source of error. Finally, we note that
apart from leakage and swap error, phase error $\delta_{\phi}$
resulted from parasitic ZZ coupling limits the performance of iSWAP gates in the
traditional AA-type setup \cite{R2,R11,R12}. The high-fidelity iSWAP gate and the low
conditional phase error $\delta_{\phi}$ demonstrated above indicate that parasitic ZZ coupling
is indeed heavily suppressed in the AB-type setup.


In summary, we have studied parasitic ZZ coupling in a superconducting architecture
\cite{R41,R42,R43} where two qubits with opposite-sign anharmonicities are coupled
together and found that high-contrast
control over parasitic ZZ coupling can be realized.
We further show that CZ gates with higher gate speed and iSWAP gates with
dramatically lower conditional phase error can be realized with diabatic schemes in
the proposed architecture. Moreover, as shown in Fig.$\,$4(b) and 4(c),
XY gates with arbitrary swap angles \cite{R44}, leakage error below $10^{-3}$,
and negligible phase error is achievable , as is arbitrary control
phase gate with swap error below $10^{-3}$. Since these errors are caused by
off-resonant Rabi oscillation related to the associated parasitic interaction
(i.e., $|01\rangle\leftrightarrow|10\rangle$ for CZ gates, and
$|11\rangle\leftrightarrow|20\rangle\,(|02\rangle)$ for iSWAP gates),
even lower error rates should be possible by increasing the value of anharmonicity \cite{R43} or
using the synchronization procedure \cite{R11}. Implementing
these continuous set of gates natively could be useful for near-term
applications of quantum processors \cite{R12,R44}. As one may expect, the high-contrast
control over ZZ coupling could also improve
the performance of parametric activated gates \cite{R15,R16,R17} and cross-resonance
gates \cite{R8,R14}. In multi-qubit systems and with fixed coupled cases, the
crosstalk resulted from ZZ coupling could be heavily suppressed, thus, gate
operations can be implemented simultaneously with low crosstalk. For tunable
coupled cases \cite{R45,R46,R47,R48}, XY gates with arbitrary swap angles can be
implemented natively with negligible conditional phase error \cite{R2,R11,R12}.

\begin{acknowledgements}
We would like to thank Yu Song for helpful suggestions
on the manuscript. This work was partly supported by
the National Key Research and Development Program of
China (Grant No. 2016YFA0301802), NSFC (Grants
No. 61521001 and No. 11890704), and the Key
R$\&$D Program of Guangdong Province (Grant
No. 2018B030326001). P. X. acknowledges the supported
by the Scientific Research Foundation of Nanjing University
of Posts and Telecommunications (NY218097), NSFC
(Grant No. 11847050), and the Young fund of
Jiangsu Natural Science Foundation of China (Grant
No. BK20180750). H. Y. acknowledges support from the
Beijing Natural Science Foundation (Grant No. Z190012).
\end{acknowledgements}

%

\clearpage




\section*{Supplementary material}

# Supplementary information for "High-contrast ZZ interaction using superconducting qubits with opposite-sign anharmonicity"

Peng Zhao,[1, *] Peng Xu,[1, 2, 3] Dong Lan,[1] Ji Chu,[1] Xinsheng Tan,[1, †] Haifeng Yu,[1] and Yang Yu[1]

[1]*National Laboratory of Solid State Microstructures, School of Physics, Nanjing University, Nanjing 230039, China*
[2]*Institute of Quantum Information and Technology, Nanjing University of Posts and Telecommunications, Nanjing, Jiangsu 210003, China*
[3]*State Key Laboratory of Quantum Optics and Devices, Shanxi University, Taiyuan, 030006, China*
(Dated: October 26, 2020)

## I. CIRCUIT HAMILTONIAN

Figure 1 shows the circuit model of a transmon qubit [1] coupled directly to a capacitively-shunted flux qubit (C-shunted flux qubit) [2] via a capacitor with the coupling capacitance $C_c$. The self-capacitance and coupling energy of the Josephson junction are denoted by $C_{J,i}$ and $E_{J,i}$, where $i = T$ and $i = 1, 2, 3$ are labels of the junctions of the transmon qubit and C-shunted flux qubit, respectively. $C_{s,T(F)}$ represents the shunted capacitor for the transmon qubit (C-shunted flux qubit), and the phase difference across the Josephson junction $i$ is denoted as $\varphi_i$. The Lagrangian of the system can be written as $\mathcal{L} = T - U$, where $T$ and $U$ denote the kinetic energy and the potential energy, respectively. We obtain [3] ($\hbar = 1$)

$$\begin{aligned}T &= \frac{1}{2}(\frac{1}{2e})^2[(C_{s,T} + C_{J,T})\dot{\varphi}_T^2 + C_c(\dot{\varphi}_T - \dot{\varphi}_3)^2 \\ &\quad + (C_{s,F} + C_{J,3})\dot{\varphi}_3^2 + C_{J,1}\dot{\varphi}_1^2 + C_{J,2}\dot{\varphi}_2^2] \\ &= \frac{1}{2}(\frac{1}{2e})^2[(C_{s,T} + C_{J,T} + C_c)\dot{\varphi}_T^2 - 2C_c\dot{\varphi}_T\dot{\varphi}_3 \\ &\quad + (C_{s,F} + C_{J,3} + C_c)\dot{\varphi}_3^2 + C_{J,1}\dot{\varphi}_1^2 + C_{J,2}\dot{\varphi}_2^2]\end{aligned} \quad (1)$$

and

$$U = E_{J,T}\cos(\varphi_T) + E_{J,1}\cos(\varphi_1) + E_{J,2}\cos(\varphi_2) \\ + E_{J,3}\cos(\varphi_3). \quad (2)$$

By choosing $\varphi_{p(m)} = (\varphi_1 \pm \varphi_2)/2$, and considering the flux quantization condition, which gives $\varphi_3 + \varphi_2 - \varphi_1 = 2\pi f$ ($f = \Phi_{\text{ext}}/\Phi_0$, where $\Phi_{\text{ext}}$ denotes the external magnetic flux threading the loop of the C-shunt flux qubit and $\Phi_0 = h/2e$ is the single flux quantum), we can write the full system Hamiltonian as

$$H = \sum_{\varphi_i} \dot{\varphi}_i \frac{\partial \mathcal{L}}{\partial \dot{\varphi}_i} - \mathcal{L} = H_T + H_F + H_c. \quad (3)$$

$H_T$ is the Hamiltonian of the transmon qubit,

$$\begin{aligned}H_T &= \frac{1}{2}(\frac{1}{2e})^2(C_{s,T} + C_{J,T} + C_c)\dot{\varphi}_T^2 - E_{J,T}\cos(\varphi_T) \\ &= \frac{1}{2}(\frac{1}{2e})^2 C_T \dot{\varphi}_T^2 - E_{TJ}\cos(\varphi_T) \\ &= 4E_{C,T} n_T^2 - E_{J,T}\cos(\varphi_T),\end{aligned} \quad (4)$$

where $E_{C,T} = e^2/2C_T$ ($C_T = C_{s,T} + C_{J,T} + C_c$) denotes the qubit charging energy ($E_{J,T}/E_{C,T} \gg 1$ for qubit working in transmon regime), and $n_T = \partial \mathcal{L}/\partial \dot{\varphi}_T = C_T \dot{\varphi}_T/(2e)^2$ is the conjugate variable of the phase differences $\varphi_T$. $H_F$ is the Hamiltonian of the C-shunted flux qubit,

$$\begin{aligned}H_F &= \frac{1}{2}(\frac{1}{2e})^2[(C_{s,F} + C_{J,3} + C_c)\dot{\varphi}_m^2 + C_{J,1}(\dot{\varphi}_p + \dot{\varphi}_m)^2 \\ &\quad + C_{J,2}(\dot{\varphi}_p - \dot{\varphi}_m)^2] - [E_{J,1}\cos(\varphi_p + \varphi_m) \\ &\quad + E_{J,2}\cos(\varphi_p - \varphi_m) + E_{J,3}\cos(2\pi f + 2\varphi_m)] \\ &= \frac{1}{2}(\frac{1}{2e})^2[(C_{s,F} + C_{J,3} + C_c + C_{J,1} + C_{J,2})\dot{\varphi}_m^2 \\ &\quad + (C_{J,1} + C_{J,2})\dot{\varphi}_p^2 + 2(C_{J,1} - C_{J,2})\dot{\varphi}_p\dot{\varphi}_m] \\ &\quad - [(E_{J,1} + E_{J,2})\cos(\varphi_p)\cos(\varphi_m) \\ &\quad + (E_{J,2} - E_{J,1})\sin(\varphi_p)\sin(\varphi_m) \\ &\quad + E_{J,3}\cos(2\pi f + 2\varphi_m)].\end{aligned} \quad (5)$$

$H_c$ is the Hamiltonian describing the inter-qubit coupling

$$H_c = -2(\frac{1}{2e})^2 C_c \dot{\varphi}_T \dot{\varphi}_m. \quad (6)$$

In this work, following the result presented in the supplementary materials of Ref. [4], we consider that the parameters of C-shunt flux qubit satisfy the following conditions, (i) the two larger junctions of the C-shunted flux qubit are identical, resulting $E_{J,1} = E_{J,2} \equiv E_J$ and $C_{J,1} = C_{J,2} \equiv C_J$; (ii) the parameters of the smaller junction is given as $E_{J,3} = \beta E_J$, $C_{J,3} = \beta C_J$, and $\beta < 0.5$, i.e., the C-shunt flux qubit operates in the single well regime [4–6]; (iii) the shunt capacitance $C_{s,F}$ is far larger than the self-capacitance of the junction, i.e., $C_{s,F} \gg C_J$. Therefore, the characteristic frequency of the

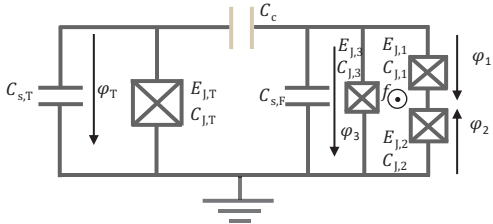

FIG. 1: Circuit model of a transmon qubit capacitively coupled to a capacitively-shunted flux qubit (C-shunt flux qubit).

p-mode ($\varphi_p$) of the C-shunt flux qubit is far larger than the typical frequency involved in present work, and can be safely omitted [4]. Then the Hamiltonian of the C-shunt flux qubit at the optimal bias point (flux insensitive point with $f = 0.5$) can be simplified as [4]

$$\begin{aligned} H_F &= \frac{1}{2}(\frac{1}{2e})^2[(C_{s,F} + \beta C_J + C_c + 2C_J)\dot\varphi_m^2 \\ &\quad - [2E_J \cos(\varphi_m) + \beta E_J \cos(2\pi f + 2\varphi_m)] \\ &= 4E_{C,m} n_m^2 - 2E_J \cos(\varphi_m) + \beta E_J \cos(2\varphi_m), \end{aligned} \quad (7)$$

where $E_{C,m} = e^2/2C_m$ ($C_m = C_{s,F} + \beta C_J + C_c + 2C_J$) is the charging energy of the m-mode ($\varphi_m$), and $n_m = \partial \mathcal{L}/\partial \dot\varphi_m = C_m \dot\varphi_m/(2e)^2$ is the conjugate variable of the phase differences $\varphi_m$. Therefore, the interaction Hamiltonian of Eq. (6) can be rewritten as

$$H_c = \frac{-2C_c(2e)^2}{C_T C_m} n_T n_m. \quad (8)$$

Following the result presented in Ref. [1] and Ref. [4], by using the cosine expansion to the leading order and introducing the annihilation and creation operators for the harmonic (linear) parts of the circuit, the Hamiltonian of the transmon qubit in Eq. (4) and C-shunt flux qubit in Eq. (7) can be approximated by [with Rotating Wave Approximation (RWA)] [1]

$$\begin{aligned} H_T &= 4E_{C,T} n_T^2 - E_{J,T} \cos(\varphi_T) \\ &= 4E_{C,T} n_T^2 - E_{J,T}[1 - \frac{\varphi_T^2}{2} + \frac{\varphi_T^4}{24} + \mathcal{O}(6)] \\ &\approx 4E_{C,T} n_T^2 + \frac{E_{J,T}}{2} \varphi_T^2 - \frac{E_{J,T}}{24} \varphi_T^4 + \text{constant} \\ &\approx \omega_T q_T^\dagger q_T + \frac{\alpha_T}{2} q_T^\dagger q_T (q_T^\dagger q_T - 1) \end{aligned} \quad (9)$$

with

$$\begin{aligned} \omega_T &= \sqrt{8E_{C,T} E_{J,T}} - E_{C,T}, \\ \alpha_T &= -E_{C,T}, \\ \varphi_T &= \left(\frac{8E_{C,T}}{E_{J,T}}\right)^{1/4} \frac{q_T + q_T^\dagger}{\sqrt{2}}, \\ n_T &= \left(\frac{8E_{C,T}}{E_{J,T}}\right)^{-1/4} \frac{q_T - q_T^\dagger}{i\sqrt{2}}, \end{aligned} \quad (10)$$

and

$$\begin{aligned} H_F &= 4E_{C,m} n_m^2 - 2E_J \cos(\varphi_m) + \beta E_J \cos(2\varphi_m) \\ &= 4E_{C,m} n_m^2 - 2E_J[1 - \frac{\varphi_m^2}{2} + \frac{\varphi_m^4}{24} + \mathcal{O}(6)] \\ &\quad + \beta E_J[1 - \frac{(2\varphi_m)^2}{2} + \frac{(2\varphi_m)^4}{24} + \mathcal{O}(6)] \\ &\approx 4E_{C,m} n_m^2 + E_J(1 - 2\beta)\varphi_m^2 \\ &\quad - E_J \frac{1 - 8\beta}{12} \varphi_m^4 + \text{constant} \\ &\approx \omega_F q_F^\dagger q_F + \frac{\alpha_F}{2} q_F^\dagger q_F (q_F^\dagger q_F - 1) \end{aligned} \quad (11)$$

with

$$\begin{aligned} \omega_F &= \sqrt{16 E_{C,m} E_J(1 - 2\beta)} + \frac{8\beta - 1}{1 - 2\beta} E_{C,m}, \\ \alpha_F &= \frac{8\beta - 1}{1 - 2\beta} E_{C,m}, \\ \varphi_m &= \left(\frac{4E_{C,m}}{E_J(1 - 2\beta)}\right)^{1/4} \frac{q_F + q_F^\dagger}{\sqrt{2}}, \\ n_m &= \left(\frac{4E_{C,m}}{E_J(1 - 2\beta)}\right)^{-1/4} \frac{q_F - q_F^\dagger}{i\sqrt{2}}. \end{aligned} \quad (12)$$

The inter-qubit coupling can be descried as

$$\begin{aligned} H_c &= g(q_T^\dagger q_F + q_T q_F^\dagger) \\ g &= \frac{-C_c(2e)^2}{C_T C_m} \left(\frac{8E_{C,T}}{E_{J,T}}\right)^{-1/4} \left(\frac{4E_{C,m}}{E_J(1 - 2\beta)}\right)^{-1/4}. \end{aligned} \quad (13)$$

Therefore, the full system Hamiltonian has following form

$$\begin{aligned} H &= \omega_T q_T^\dagger q_T + \frac{\alpha_T}{2} q_T^\dagger q_T (q_T^\dagger q_T - 1) \\ &\quad + \omega_F q_F^\dagger q_F + \frac{\alpha_F}{2} q_F^\dagger q_F (q_F^\dagger q_F - 1) \\ &\quad + g(q_T^\dagger q_F + q_T q_F^\dagger), \end{aligned} \quad (14)$$

recovering the full system Hamiltonian $H$ in the main text.

From the above discussion, one can find that the C-shunted flux qubit can be approximately modeled as an anharmonic oscillator with positive anharmonicity under the condition $0.125 < \beta < 0.5$, which has been experimentally demonstrated in previous works [4–6]. More importantly, similar to that of the transmon qubit, the anharmonicity of the C-shunted flux qubit depends only on the characteristic geometry of the qubit (according to the derived expression for $\alpha_F$ in Eq. (12) for the C-shunted flux qubit), such as $\beta$ and the charge energy $E_{C,m}$ determined dominantly by the shunted capacitance $C_{s,F}$. Therefore, the reproducibility of the fabrication of C-shunted flux qubits is comparable to that of transmon qubits. The anharmonicity precision of $\sigma_\alpha < 10\,\text{MHz}$ could be achieved for the two types qubits with current fabrication techniques, as shown in Ref. [4] and Ref. [7, 8]. Finally, we note that the anharmonicity of the transmon qubit is around $-400 \sim -180\,\text{MHz}$, which have been demonstrated experimentally in various work [7–13], while the reported value of the anharmonicity is around $200 \sim 900\,\text{MHz}$ for the C-shunted flux qubit (in single well regime) [4–6].

We note that in the above discussion, the condition (i) is proposed for simplifying the analysis of the C-shunt flux qubit. However, with current junction fabrication techniques, the typical junction asymmetries $d = (E_{J,1} - E_{J,2})/(E_{J,1} + E_{J,2})$ is about 10% (one may estimate that the self-capacitance asymmetries should also have a comparable magnitude) [1], leading to the inter-mode coupling presented in Eq. (5). Fortunately, with typical parameters, the characteristic frequency of the p-mode ($\varphi_p$) of a C-shunt flux qubit is far larger than the typical frequency (such as the $m$-mode

transition frequency and transmon qubit frequency) involved in present work (Typically, 50 GHz vs 5 GHz), thus in principle can be safely omitted [4]. To the best of our knowledge, for the C-shunt flux qubit, there are no experimental result that has shown the detrimental effect resulted from the junction asymmetry [4–6]. In fact, previous experimental result has shown that the C-shunt flux qubit (in the single well regime, $0.125 < \beta < 0.5$) can have a coherence time comparable to that of the transmon qubit [4].

Overall, one can reasonably estimate that the C-shunt flux qubit (in single well regime) can be modeled as an anharmonic oscillator with positive anharmonicity, and the magnitude of the qubit anharmonicity that is comparable to that of the transmon qubit, should be achievable with current fabrication techniques [4–6].

## II. TRIPLE DEGENERACY POINT

As shown in the inset of Fig. 2(b) in the main text, for architecture consisting of two direct-coupled qubits with opposite-sign anharmonicities, the interaction among the two-excitation manifold consisting of qubit state $|11\rangle$ and the non-qubit states ($|20\rangle$, $|02\rangle$) forms a triple degeneracy point, when the qubit detuning equals the value of the anharmonicity of qubits ($\Delta = \alpha_b$). Here, we give a detailed description of interaction among this two-excitation manifold (As same as the discussion in the main text, the frequency of qubit $b$ is fixed). By assuming that the constant energy of non-qubit state $|02\rangle$ is zero, i.e., $E_{02} = 0$ ($E_{ij}$ is an eigenenergy of the non-interacting Hamiltonian, i.e., $H$ with $g = 0$, and the corresponding eigenstate is bare state $|ij\rangle$), the Hamiltonian of the system truncated to the two-excitation manifold is

$$H_{\text{tri}} = \begin{pmatrix} 2\delta & J & 0 \\ J & \delta & J \\ 0 & J & 0 \end{pmatrix} \quad (15)$$

where $J = \sqrt{2}g$ is the coupling strength between $|11\rangle$ and non-qubit states ($|02\rangle$, $|20\rangle$), and $\delta = \Delta - \alpha$. By defining $\theta = \arctan[\delta/(\sqrt{2}J)]$, the eigenstates of this Hamiltonian are

$$\begin{aligned}|\psi\rangle_1 &= \frac{[(1+\sin\theta)|02\rangle + (1-\sin\theta)|20\rangle - \sqrt{2}\cos\theta|11\rangle]}{2} \\ |\psi\rangle_2 &= \frac{[\cos\theta(|02\rangle - |20\rangle) + \sqrt{2}\sin\theta|11\rangle]}{\sqrt{2}} \\ |\psi\rangle_3 &= \frac{[(1-\sin\theta)|02\rangle + (1+\sin\theta)|20\rangle + \sqrt{2}\cos\theta|11\rangle]}{2}\end{aligned} \quad (16)$$

with the corresponding energies of $\delta - \sqrt{2J^2 + \delta^2}$, $\delta$, and $\delta + \sqrt{2J^2 + \delta^2}$. At the triple degeneracy point where the qubit detuning equals the value of the anharmonicity of qubits, i,e., $\delta = \Delta - \alpha = 0$, the three eigenstates are $(|02\rangle + |20\rangle - \sqrt{2}|11\rangle)/2$, $(|02\rangle - |20\rangle)/\sqrt{2}$, $(|02\rangle + |20\rangle + \sqrt{2}|11\rangle)/2$, with the corresponding energies of $-\sqrt{2}J$, $0$, and $\sqrt{2}J$.

From the above discussion and also the result shown in Fig. 3(a) of the main text, where an unphysical sudden change of the ZZ coupling strength has presented for system approaching the triple degeneracy point, the ZZ coupling (according to the definition in Eq. (2) of the main text) for AB-type setup is not well defined for region close to the triple degeneracy point. This is caused by the fact that the interaction among two-excitation manifold hybridize the bare qubit states $|11\rangle$ and bare non-qubit state $\{|02\rangle, |20\rangle\}$, forming the eignstates described by Eq. (16). Thus, from Eq. (16), one can find that in order to calculate the ZZ coupling strength, one needs to decide how to extend the state labeling scheme adopted in the regime away from the triple degeneracy point into regions where there are avoided crossings with triples. For the case of the AA-type setup, the labeling scheme defined in main text, i.e., the state $|\widetilde{ij}\rangle$ denotes the eigenstate of the Hamiltonian $H$ that has the maximum overlap with the bare state $|ij\rangle$, makes sense in the whole regime including the degeneracy points at $\Delta = \pm\alpha$, as shown in Fig. 3(a) of the main text, and it is compatible with the point view of adiabatic theorem, i.e., the state $|\widetilde{ij}\rangle$ is adiabatically connected to the bare state $|ij\rangle$. For the case of AB-type setup, the labeling scheme defined in main text is reasonable in the regime away from the triple degeneracy point ($|\langle 11|\psi_2\rangle|^2 > |\langle 11|\psi_{1,3}\rangle|^2$), i.e., since the eigenstate $|\psi_2\rangle$ has the maximum overlap with the bare state $|11\rangle$, $\psi_2$ is chosen as $|\widetilde{11}\rangle$ for calculating the ZZ coupling strength according to Eq. (2) in the main text, and its result is also compatible with the point view of adiabatic theorem. However, for the regime close to the triple degeneracy point ($|\langle 11|\psi_2\rangle|^2 < |\langle 11|\psi_{1,3}\rangle|^2$), the present labeling scheme labels $|\psi\rangle_{1,3}$ as the $|\widetilde{11}\rangle$ in calculating the ZZ coupling strength, causing the unphysical sudden change of the strength, as shown in Fig. 3(a) of the main text, and its result is not compatible with the point view of adiabatic theorem. In fact, from the point view of adiabatic theorem, i.e., the labeled $|\widetilde{ij}\rangle$ should be adiabatically connected to the bare state $|ij\rangle$, and in this setting, the ZZ coupling strength is zero in the whole regime including the triple degeneracy point, thus the CZ gate operation is infeasible for the adiabatic scheme, i.e., by adiabatically varying the qubit frequency into the interaction point.

In present work, we choose the definition given in Eq. (2) of the main text to calculate the ZZ coupling strength in the whole regime due to the two benefits for analyzing the ZZ coupling : (i) as we have mentioned above, the CZ gate operation in present system with zero anharmonicity asymmetry is infeasible for the adiabatic scheme (Note that even for system with none-zero anharmonicity asymmetry, the ZZ coupling is heavily suppressed as compared with that of the AA-type setup, as shown in Fig. 3(b) of the main text, thus the gate speed is slower for the adiabatic scheme). Hence, in present work, we utilize the diabatic scheme where a full swap between qubit state $|11\rangle$ and non-qubit states is required for implementing CZ gate. By using the labeling scheme defined in the main text, the calculated result presented in Fig. 3(a), gives

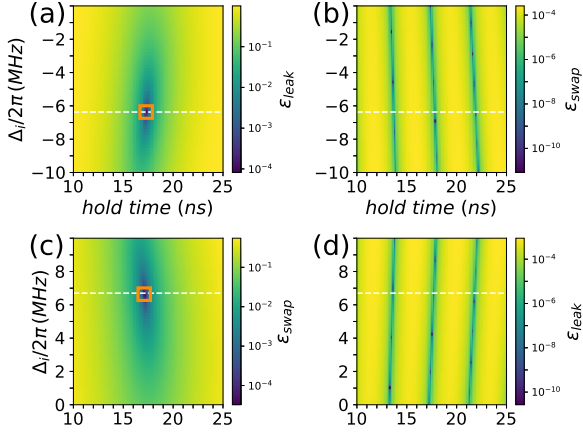

FIG. 2: (a),(b) Swap error ($\varepsilon_{\text{swap}} = 1 - P_{\overline{01}}$) and leakage ($\varepsilon_{\text{leak}} = 1 - P_{\overline{11}}$) during the CZ gate operation with diabatic scheme as a function of hold time and frequency overshot ($\Delta_i$) for system initiated in $|\overline{11}\rangle$ and $|\overline{01}\rangle$ at the parking point, respectively. (c),(d) Leakage error ($\varepsilon_{\text{leak}} = 1 - P_{\overline{11}}$) and swap error ($\varepsilon_{\text{swap}} = P_{\overline{01}}$) during the iSWAP gate operation with diabatic scheme as a function of hold time and frequency overshot ($\Delta_i$) for system initiated in $|\overline{11}\rangle$ and $|\overline{01}\rangle$ at the parking point, respectively. The horizontal cuts (dashed lines) depict the optimal value of overshot used in the main text. The orange squares indicate the 'optimal' point used in the main text for the implementation of iSWAP gate and CZ gate

the swap rate ($\sqrt{2}J = 2g$, i.e., the ZZ coupling strength at the triple degeneracy point) between the qubit state and non-qubit state. Thus, we can find that in the AB-setup, the CZ gate can be realized with gate speed faster than that of the AA-setup, where the swap rate is $J = \sqrt{2}g$ at the degeneracy point. (ii) the definition of taken in present work is convenient to numerically analysis the ZZ coupling in the whole parameters space.

## III. GATE OPERATION WITH DIABATIC SCHEME

As mentioned in the main text, we consider implementing the CZ gate and iSWAP gate by using diabatic scheme in our proposed system comprising two qubits with oposite-sign anharmonicities. During the gate implementation, the frequency of qubit $b$ keeps at its parking point, while the frequency of qubit $a$ moves from its parking point to the interaction point, and then comes back according to a time-dependent function (the rounded trapezoid-shaped pulses), given as [14]

$$\omega_a(t) = \omega_P + \frac{\omega_I - \omega_P}{2}[\text{Erf}(\frac{t - \frac{1}{2}t_{\text{ramp}}}{\sqrt{2}\sigma}) - \text{Erf}(\frac{t - t_{\text{gate}} + \frac{1}{2}t_{\text{ramp}}}{\sqrt{2}\sigma})] \quad (17)$$

where $\omega_P$ and $\omega_I$ denote the qubit frequency at the parking point (the logical states are defined as the eigenstates of the system Hamiltonian biased at this point) and the interaction point, respectively. $t_{\text{ramp}} = 4\sqrt{2}\sigma$ ($\sigma = 1\,\text{ns}$) is the ramp time, and $t_{\text{gate}}$ denotes the total gate time for implementing the two-gate operations. $t_{\text{hold}} = t_{\text{gate}} - t_{\text{ramp}}$ denotes the hold time that is defined as the time-interval between the midpoints of the ramps, as shown in Fig. 4(a) and 4(b) of the main text. In this way, by assuming no leakage to non-qubit states and excluding the decoherence process, up to single qubit phase gates, the actually implemented unitary gate can be described by

$$U(\theta, \phi) = \begin{pmatrix} 1 & 0 & 0 & 0 \\ 0 & \cos(\theta) & -i\sin(\theta) & 0 \\ 0 & -i\sin(\theta) & \cos(\theta) & 0 \\ 0 & 0 & 0 & e^{-i\phi} \end{pmatrix} \quad (18)$$

where $\theta$ represents the swap angle associated with the bare exchange interaction between $|10\rangle$ and $|01\rangle$, and $\phi$ is the conditional phase resulting from the parasitic ZZ coupling. An ideal iSWAP (CZ) gate can be described by $U(\pi/2, 0)$ ($U(0, \pi)$).

In order to quantify the intrinsic performance (excluding the decoherence error) of the implemented two-qubit gates, we use the metric of state-average gate fidelity in present work. For two-qubit gate, the fidelity is defined as [15]

$$F \equiv \frac{\text{Tr}(U^\dagger U) + |\text{Tr}(U_{\text{target}}^\dagger U)|^2}{20} \quad (19)$$

where $U$ is the actual evolution operator after applying an auxiliary single-qubit $Z$ rotation on each of the two-qubits before and after the gate implementation, and truncated to the computational subspace (spanned by the logical eigenstates at the parking point) [14, 16, 17], and $U_{\text{target}}$ denotes the target gate operations.

For the system with Hamiltonian $H$ and the control pulse of Eq. (17), the actual evolution operator in the rotating frame with respect to the Hamiltonian at parking point (with $H(0)$) is given as

$$U_{sys} = \hat{\mathcal{T}} \exp\left(-i \int_0^{t_{\text{gate}}} H_R(t) dt\right), \quad (20)$$

where $H_R(t) = e^{iH(0)t}H(t)e^{-iH(0)t} - H(0)$, and $\hat{\mathcal{T}}$ denotes the time-ordering operator. Thus, $U$ in Eq. (19) is given as

$$U = U_{\text{post}} \mathcal{P} U_{sys} \mathcal{P}^\dagger U_{\text{pre}}, \quad (21)$$

where $\mathcal{P}$ represents the projected operator defined in the computational subspace of the full syatem, and $U_{\text{post}}$ and $U_{\text{pre}}$ are the auxiliary single-qubit $Z$ rotations on the two qubit.

$$U_{\text{post}} = e^{-i\phi_1 ZI/2}e^{-i\phi_2 IZ/2}, U_{\text{pre}} = e^{-i\phi'_1 ZI/2}e^{-i\phi'_2 IZ/2}, \quad (22)$$

where $Z$ and $I$ denote the Pauli operator and identity operator (defined in the computational subspace of the full system), respectively. Therefore, in present work, the fidelity of the two-qubit gate is obtained as

$$F = \text{maximize}_{\phi_j, \phi'_j} F(\phi_j, \phi'_j). \quad (23)$$

4

As we discussed above, when the leakage error is so small that can be ignored, the actually implemented gate can be approximated by $U(\theta, \phi)$ in Eq. (18). Thus, one can extract the phase parameters $(\theta, \phi)$ as

$$\{\theta, \phi\} = \arg[\text{maximize}_{\phi_j, \phi_j', \{\theta, \phi\}} \ F(\phi_j, \phi_j', \{\theta, \phi\})]. \quad (24)$$

where $F(\phi_j, \phi_j', \{\theta, \phi\})$ is gate fidelity with respect to the target gate operation $U(\theta, \phi)$ given in Eq. (18) with two free parameters: $\theta$ and $\phi$.

### A. Optimal control parameters

In our system, the qubit coupling strength is far smaller than the qubit frequencies (typically, $g/\omega_q < 0.003$), the excitation non-conservation process can be safely omitted, i.e., the Rotating Wave Approximation (RWA) is valid. Thus, the swap error is only caused by unintended swap process between $|01\rangle$ and $|10\rangle$, and the leakage error is only caused by the unintended swap process between $|11\rangle$ and $\{|02\rangle, |20\rangle\}$. Therefore, in present work, the metric $\varepsilon_{\text{swap}} = 1 - P_{\bar{0}1}$ (or $\varepsilon_{\text{swap}} = P_{\bar{0}1}$) and $\varepsilon_{\text{leak}} = 1 - P_{\bar{1}1}$ are used to quantify the swap error and leakage error, respectively. From the expression of Eq. (17), to minimize the swap error and leakage error during the implementation of the iSWAP gate and CZ gate, we minimize these errors by optimizing the control parameters: the hold time and a smaller frequency overshot $\Delta_i$ relative to the interaction frequency $\omega_I$.

As discussed in Ref. [18], and also shown in Fig. 2(a) and 2(c), the overshoot $\Delta_i$ is critical to optimize the leakage error and swap error for the implementation of the CZ gate and iSWAP gate, respectively, while the swap error for CZ gate and the leakage error for iSWAP gate are insensitive to this small frequency overshoot, as shown in Fig. 2(b) and 2(d). The dashed line shows the optimal value of the overshoot adopted in the main text for the implementation of CZ gate and iSWAP gate. However, from the result shown in Fig. (2), one can find that in the present system with fixed inter-qubit coupling, it is nearly impossible to have an optimal hold time for simultaneously minimizing the swap error and leakage error. Meanwhile, one can find that the swap error for CZ gate operation and the leakage error for iSWAP gate are both suppressed below $10^{-3}$, as shown in Figs. 4(b) and 4(e) of the main text. In fact, since these errors are caused by the off-resonant Rabi oscillation related to the associated parasitic interactions (i.e., $01 \leftrightarrow 10$ for CZ gate operations, and $11 \leftrightarrow 20 \ (02)$ for iSWAP gate operations), even lower error should be possible by increasing the value of qubit anharmonicity [19] or applying the procedure of synchronization introduced in Ref. [18]. Thus, in the present work, the optimal hold time is chosen to minimize the leakage error (swap error) for CZ (iSWAP gate), as shown in Fig. 2(a) and 2(c).

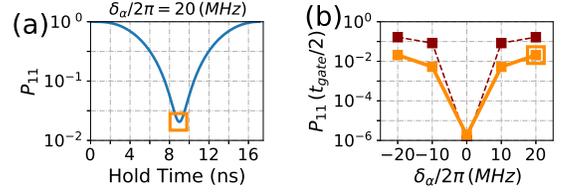

FIG. 3: (a) Population $P_{\bar{1}1}$ versus hold time for system initiated in $|\overline{11}\rangle$ at the parking point. The anharmonicity asymmetry is $\delta_\alpha/2\pi = 20\,\text{MHz}$. The orange square indicates the population in $|\overline{11}\rangle$ at time $t_{\text{gate}}/2$. (b) Population $P_{\bar{1}1}(t_{\text{gate}}/2)$ versus anharmonicity asymmetry $\delta_\alpha$. The phase error $\delta_\phi$ is also displayed (dashed line) for easy reference.

### B. CZ gate operation

As shown in Fig. 3(c) of the main text, in order to implement the CZ gate in our proposed system, we apply the rounded trapezoid-shaped pulses described by Eq. (17) to adjust the frequency of qubit $a$ from the parking point $\omega_P/2\pi \simeq 6.1\,\text{GHz}$ to the interaction point $\omega_I = \omega_b + \alpha$, and the freqiency of qubit $b$ keeps fixed. The optimal working point is indicated in Fig. 2(a), where hold time is $t_{\text{hold}} = 17.3\,\text{ns}$ and overshoot is $\Delta_i/2\pi = -6.37\,\text{MHz}$. According to Eq. (19) with $U_{\text{target}} \equiv U(0, \pi)$, the intrinsic gate fidelity is about 99.999%. The leakage error $\epsilon_{\text{leak}}$ is below $7.73 \times 10^{-7}$. Thus, the leakage error can be safely omitted. Using Eq. (24), we can approximately obtain the phase parameters of the implemented gate $U(\theta = 0 + \delta_\theta, \phi = \pi + \delta_\phi)$, from which we find that the phase error $\delta_\theta/\pi = -1.167 \times 10^{-3}$ and $\delta_\phi/\pi$ is lower than $10^{-7}$.

The above results are obtained for anharmonicity asymmetry $\delta_\alpha = 0$, i.e., the two qubits have exactly opposite anharmonicities. However, as shown in Fig. 3(c) of the main text, in a practical system with typical anharmonicity asymmetry $\delta_\alpha$ around $-20 \sim 20\,\text{MHz}$, the fidelity of the implemented CZ gates get worse for system with none-zero anharmonicity asymmetry $\delta_\alpha$. By extracting the phase error $(\delta_\theta, \delta_\phi)$ and the leakage error $\varepsilon_{\text{leak}}$, one can find that for implementing CZ gate in system with large anharmonicity asymmetry, the dominated error source is the phase error $\delta_\phi$, as shown in Fig. 3(c) of the main text. As discussed in the main text, the implementation of the CZ gate requires a full Rabi oscillation between qubit state $|11\rangle$ and non-qubit states, and in a system with zero anharmonicity asymmetry, at the triple degeneracy point, the interaction among the two-excitation manifold can cause a full Rabi oscillation between $|11\rangle$ and $[|02\rangle + |20\rangle]/\sqrt{2}$, thus accumulating a conditioned phase $\pi$. However, in a system with none-zero anharmonicity asymmetry, the resonance condition breaks down, hence an off-resonance Rabi oscillations between the qubit state $|11\rangle$ and non-qubit states ($|02\rangle$, $|20\rangle$) [18] presents, as shown in Fig. 3. Therefore, the accumulated conditional phase is not equals to $\pi$.

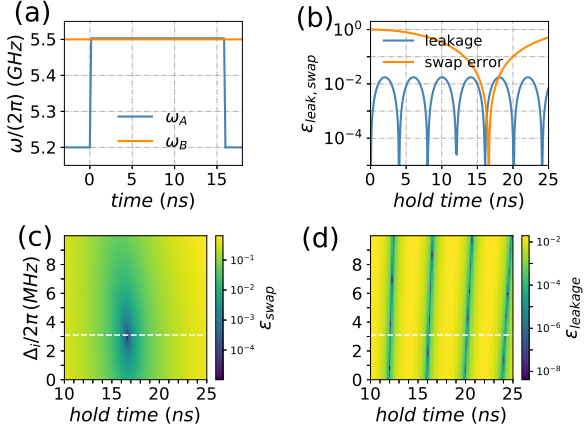

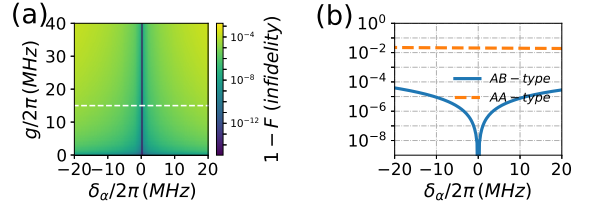

FIG. 4: Numerical result for implementing iSWAP gate with a rectangle pulse. (a) Rectangle pulse with small overshot for realizing iSWAP gate in the proposed architecture, and the system parameters are the same as in Fig. 4(d). (b) Leakage and swap error versus hold time. (c),(d) Swap error ($\varepsilon_{\rm swap} = P_{0\bar{1}}$) and leakage ($\varepsilon_{\rm leak} = 1-P_{\bar{1}\bar{1}}$) as function of hold time and frequency overshoot ($\Delta_i$) for system initiated in $|\bar{1}\bar{1}\rangle$ and $|\bar{0}\bar{1}\rangle$ at the parking point, respectively. The horizontal cuts (dashed lines) depict the optimal value of overshot used in (b).

### C. iSWAP gate operation

As discussed in the main text, apart from the leakage error and control error, the coherent phase error resulting from the parasitic ZZ coupling now limits the performance of fast iSWAP gate in traditional setup [8, 18, 20]. Here, we give a rough estimation of the phase error by anharmonicity asymmetry during the implementation of iSWAP gate as a function of coupling strength $g$ in our architecture. As shown in Fig. 4(a), we consider that the frequency of qubit $a$ varies from the parking point to the interaction point following a rectangle pulse, and the frequency of qubit $b$ is fixed. As shown in Fig. 4(b), for the rectangle pulse, the leakage error and swap error can also be suppressed below $10^{-3}$, and should be further reduced below $10^{-5}$ with the procedure of synchronization introduced in Ref. [18].

In order to quantify the gate infidelity caused by the phase error, here we consider that the leakage error and swap error is so small that can be ignored. The only error source is the phase error. Therefore, we can use the metric $1 - F$ where $F = [\text{Tr}(U^\dagger U) + |\text{Tr}(U_{\rm ideal}^\dagger U)|^2]/20 = [4 + |3 + e^{-i\phi}|^2]/20$ is the average gate fidelity, to quantify the phase error. For implementing the iSWAP gate with a rectangle pulse, the accumulated phase during the gate operations is $\phi = \zeta_I t_{\rm gate} = \pi\zeta_I/2g$, where $\zeta_I$ denotes the parasitic ZZ coupling strength at the interaction point, and $t_{\rm gate}$ is the gate time. Fig. 5(a) shows the result of the gate infidelity as a function of the coupling strength $g$ and anharmonicity asymmetry $\delta_\alpha$. For system with a typical coupling strength $g/2\pi$ of 15 MHz, one can find that the infidelity is suppressed below $10^{-4}$ for typical anharmonicity asymmetry around $-20 \sim 20\,{\rm MHz}$, while for the traditional setup, the gate fidelity is limited to less than 99%.

## IV. QUBITS COUPLED VIA A COUPLER

In principle, the two qubits can be coupled directly, the spectrum of which is shown in Fig. 2 in the main text, and they can also be indirect-coupled via a coupler. Typically, the coupler circuit can be a resonator with or without anharmonicity (Kerr interaction), a tunable inductor [1], or an effective tunable capacitor [9, 22, 23] such as the tunable coupler circuits combining a capacitor and a bus resonator [22, 23]. In the following, we give a detailed analysis of the system with coupler using linear resonator and an effective tunable capacitor.

### A. Resonator

For two qubits coupled via a resonator, the Hamiltonian of the system is given as

$$H = \left[\sum_{l=a,b,c} \omega_l q_l^\dagger q_l + \frac{\alpha_l}{2} q_l^\dagger q_l(q_l^\dagger q_l - 1)\right] + \sum_{l=a,b}\left[g_l(q_c^\dagger q_l + q_c q_l^\dagger)\right], \quad (25)$$

where the subscript $l = a, b, c$ labels different-type anharmonic oscillator with anharmonicity $\alpha_l$ and frequency $\omega_l$, $g_l$ denotes the coupling strength between oscillators, and $q_l$ ($q_l^\dagger$) is the associated annihilation (creation) operator truncated to the lowest three-level.

Here, we consider that subscript $l = a, b$ labels the two qubits, $\Delta = \omega_a - \omega_b$ denotes the qubit detuning, and $l = c$ labels a bus resonator with anharmonicity of $\alpha_c = 0$. For illustration purposes, we use the following parameters: qubit $b$ frequency $\omega_b/2\pi = 4.914\,{\rm GHz}$, resonator frequency $\omega_c/2\pi = 6.31\,{\rm GHz}$, value of anharmonicity $\alpha/2\pi = 330\,{\rm MHz}$, and qubit-resonator coupling strength $g_{a(b)}/2\pi = 138(135)\,{\rm MHz}$.

FIG. 5: (a) Phase error caused by the parasitic ZZ coupling during iSWAP gate operations versus coupling strength $g$ and anharmonicity asymmetry $\delta_\alpha$. (b) Phase error versus anharmonicity asymmetry for $g/2\pi = 15\,{\rm MHz}$, where the solid line (the horizontal cuts through (a)) is for our architecture, and the dashed line is for AA-type setup.


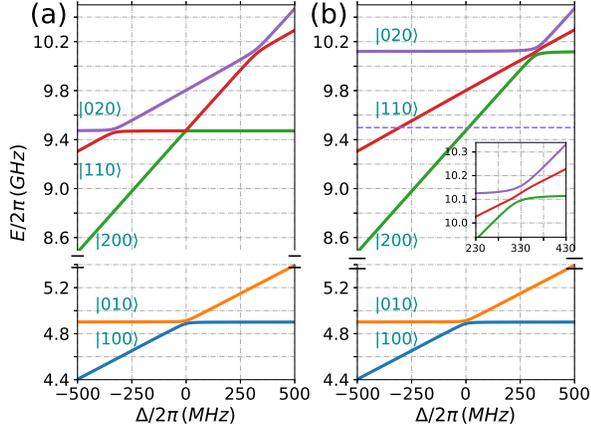

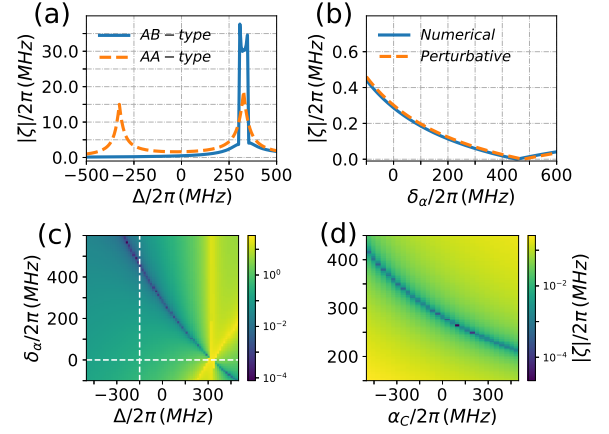

FIG. 6: Numerical calculation of the energy levels of coupled qubit system, where qubits are coupled via a resonator, as a function of the qubit detuning $\Delta = \omega_a - \omega_b$. (a) Energy levels of coupled qubits with same-sign anharmonicity $\alpha_{a,b} = -\alpha$ ($\alpha/2\pi = 330$ MHz). (b) Energy levels of coupled qubits with opposite-sign anharmonicity, i.e., $\alpha_a = -\alpha$, $\alpha_b = \alpha$. The inset shows the avoided crossing mainly resulting from the interaction among $|020\rangle$, $|200\rangle$ and $|110\rangle$ (For $|ijk\rangle$, where the first two label two-qubit states, and the third one denotes state of the resonator).

FIG. 7: Numerical result for ZZ coupling strength $|\zeta|$ in the proposed architecture (AB-type), where two qubits are coupled via a bus resonator. (a)$|\zeta|$ versus qubit detuning $\Delta$ for anharmonicity asymmetry $\delta_\alpha = 0$, where the dashed line is for traditional setup (AA-type). (b) $|\zeta|$ versus $\delta_\alpha$ for $\Delta/2\pi = -150$ MHz. (c) The ZZ coupling strength $|\zeta|$ versus qubit detuning $\Delta$ and anharmonicity asymmetry $\delta_\alpha$ for $\alpha_c/2\pi = -100$ MHz. Horizontal (vertical) cuts through (c) correspond to the result shown in (a) and (b), respectively. (d) The ZZ coupling strength $|\zeta|$ versus coupler anharmonicity $\alpha_c$ and anharmonicity asymmetry $\delta_\alpha$ for $\Delta = 0$.

Fig. 6 shows the numerically calculated energy level of coupled system, which is similar to the result for direct-coupled case. However, we note that for the indirect-coupled case, the avoided crossing with triplets shown in the inset of Fig. 6(b) results from the interaction among two-excitation manifold consisting of six states, i.e., $|020\rangle$, $|200\rangle$, $|110\rangle$, $|101\rangle$, $|002\rangle$, and $|011\rangle$ (For $|ijk\rangle$, where the first two label two-qubit states, and the third one denotes state of the resonator), rather than three states in the direct-coupled case. For qubit with weak anharmonocity coupled to a resonator in dispersive regime, i.e., $|\Delta_{a(b)}| \gg g_{a(b)}$ ($\Delta_{a(b)} = \omega_{a(b)} - \omega_c$ are the qubit-resonator detuning), these states can be grouped into two distinct subsets, one with ($|020\rangle$, $|200\rangle$, and $|110\rangle$) denotes two-excitation space of qubits at an energy scale of $2\omega$ ($\omega$ denotes a typical qubit frequency), and the other with ($|101\rangle$, $|002\rangle$, and $|011\rangle$) at an energy scale of $\{\omega+\omega_c, 2\omega_c\}$ mainly depends on the resonator frequency. In the dispersive regime, the two subsets are detuned on the order of $\{\Delta_{a,b}, 2\Delta_{a,b}\}$, which is assumed to be larger than the coupling strength between the two subsets. Thus, the avoided crossing with triplets shown in the inset of Fig. 6(b) can be approximately described by the result given in Eqs. (15) and (16), i.e., the avoided crossing with triplets results from the coupling among the two-excitation subspace of qubits ($|020\rangle$, $|200\rangle$, and $|110\rangle$).

In Fig. 7, we also numerically analyze the ZZ coupling in the proposed architecture, where qubits are coupled via a resonator, and the system is described by the Hamiltonian in Eq. (25). As shown in Fig. 7(a), compared with the direct-coupled case, the ZZ coupling is not fully eliminated for $\delta_\alpha = 0$, but still heavily suppressed as compared with that of the traditional setup. Fig. 7(b) shows ZZ coupling strength as a function of anharmonicity asymmetry $\delta_\alpha$ for $\Delta/2\pi = -150$ MHz, and we have found that the zero ZZ coupling point is at about $\delta_\alpha/2\pi = 500$ MHz rather than 0 as in the direct-coupled case. This characteristic is more prominent in the full parameter space, as shown in Fig. 7(c). The physics behind this characteristic is that since the two qubits are coupled via a resonator, the effective coupling strength between the two qubits, i.e, the strength of the interaction among the higher energy levels of qubits, depends on the qubit detuning. Moreover, the higher energy level of the resonator also contributes to the ZZ coupling. To easily identify the contribution sources to the ZZ coupling, we assume that the resonator has a nonzero anharmonicity $\alpha_c$. Using perturbation theory, the fourth-order result for the ZZ interaction strength in this case is [24–26]

$$\zeta = 2g_a^2 g_b^2 \left[ \frac{1}{\Delta_a^2(-\alpha_b + \Delta)} + \frac{1}{\Delta_b^2(-\alpha_a - \Delta)} \right] \\ + \frac{2g_a^2 g_b^2}{\Delta_a + \Delta_b - \alpha_c} \left[ \frac{1}{\Delta_a} + \frac{1}{\Delta_b} \right]^2, \quad (26)$$

From Eq. (26), one can find that the first term corresponds to the contributions related to qubit anharmonicity, thus resulting from the interactions among higher energy levels of qubits, while the second one only involves bus anharmonicity $\alpha_c$, thus resulting from the interaction between the higher energy levels of the resonator and qubit state $|11\rangle$. Consequently, the zero ZZ coupling point depends not only on the anharmonicity asymmetry and the qubit detuning, but also on the anhar-

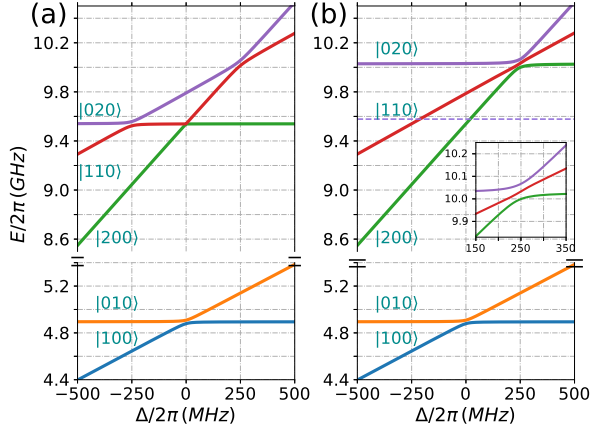

FIG. 8: Numerical calculation of the energy levels of coupled system, where two qubits are coupled via a tunable coupler, as a function of the qubit detuning $\Delta = \omega_a - \omega_b$. (a) Energy levels of coupled qubits with same-sign anharmonicity $\alpha_{a,b} = -\alpha$ ($\alpha/2\pi = 250\,\text{MHz}$). (b) Energy levels of coupled qubits with opposite-sign anharmonicity, i.e., $\alpha_a = -\alpha$, $\alpha_b = \alpha$. The inset shows the avoided crossing mainly resulting from the interaction among $|020\rangle$, $|200\rangle$ and $|110\rangle$.

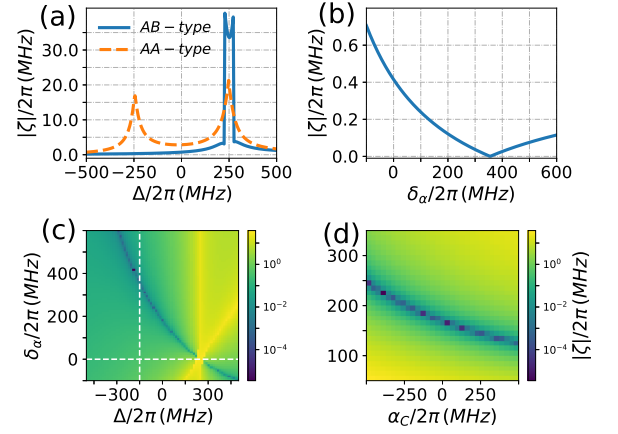

FIG. 9: Numerical result for ZZ coupling strength $|\zeta|$ in the proposed architecture with a tunable coupler (AB-type). (a) $|\zeta|$ versus qubit detuning $\Delta$ for anharmonicity asymmetry $\delta_\alpha = 0$, where the dashed line is for the traditional setup (AA-type). (b) $|\zeta|$ versus $\delta_\alpha$ for $\Delta/2\pi = -150\,\text{MHz}$. (c) The ZZ coupling strength $|\zeta|$ versus qubit detuning $\Delta$ and anharmonicity asymmetry $\delta_\alpha$ for $\alpha_c/2\pi = -100\,\text{MHz}$. Horizontal (vertical) cuts through (c) correspond to the result shown in (a) and (b), respectively. (d) The ZZ coupling strength $|\zeta|$ versus coupler anharmonicity $\alpha_c$ and anharmonicity asymmetry $\delta_\alpha$ for $\Delta = 0$.

monicity of the bus resonator. Shown in Fig. 7(d) is the ZZ coupling strength $|\zeta|$ as a function of the coupler anharmonicity $\alpha_c$ and anharmonicity asymmetry $\delta_\alpha$ for $\Delta = 0$. The characteristic provided by this coupler circuit enables us to exploit a larger parameter space for engineering the ZZ coupling.

### B. Tunable coupler

For two qubits coupled via an effective tunable capacitor combing a capacitor and a resonator, the Hamiltonian of the system is given as [23]

$$H = \left[\sum_{l=a,b,c} \omega_l q_l^\dagger q_l + \frac{\alpha_l}{2} q_l^\dagger q_l(q_l^\dagger q_l - 1)\right] \\ + \sum_{l=a,b}\left[g_l(q_c^\dagger q_l + q_c q_l^\dagger)\right] + g(q_a^\dagger q_b + q_a q_b^\dagger), \quad (27)$$

where $g$ denotes the coupling strength between the two qubits via a capacitor. The system parameters used in the following discussion are: $Q_b$'s frequency $\omega_b/2\pi = 4.914\,\text{GHz}$, resonator frequency $\omega_c/2\pi = 6.514\,\text{GHz}$ with anharmonicity $\alpha_c/2\pi = -100\,\text{MHz}$, value of qubit anharmonicity $\alpha/2\pi = 250\,\text{MHz}$, directed coupling strength $g/2\pi = 5\,\text{MHz}$ and qubit-resonator coupling strength $g_{a(b)}/2\pi = 185(176)\,\text{MHz}$.

Fig. 8 shows the numerically calculated energy level of coupled system, which is similar to the result for the direct-coupled case in the main text. Similar to the analysis of the avoided crossing shown in the inset of Fig. 7(b) for qubits coupled via a resonator, strictly speaking, the avoided crossing with triplets shown in the inset of Fig. 8(b) also results from interaction among two-excitation manifold with six states. For qubit-resonator system operated in dispersive regime, the avoided crossing can be approximately described by the result given in Eqs. (15) and (16).

In Fig. 9, we also show the numerical result of the ZZ coupling strength as a function of qubit detuning $\Delta$ and anharmonicity asymmetry $\delta_\alpha$ in this case. The result for the traditional setup is also shown for easy comparison. As shown in Fig. 9(a), the numerical result is similar to that of the resonator case shown in Fig. 7. Since parts of the net two-qubit coupling are resulting from the resonator-mediated interaction, the net effective coupling strength between the two qubits, e.g., the strength of the interaction among the higher energy levels of qubits, depends on the qubit detuning. Meanwhile, the higher energy levels of the resonator also contributes to the ZZ coupling between the two qubits. Consequently, similar to the resonator case, the zero ZZ coupling point here depends not only on the anharmonicity asymmetry, but also on the coupler anharmonicity $\alpha_c$, as shown in Fig. 9(d), where the Z-Z coupling strength $|\zeta|$ versus coupler anharmonicity $\alpha_c$ and anharmonicity asymmetry $\delta_\alpha$ is plotted for $\Delta = 0$.

---

[1] J. Koch, T. M. Yu, J. Gambetta, A. A. Houck, D. I. Schuster, J. Majer, A. Blais, M. H. Devoret, S. M. Girvin, and R. J. Schoelkopf, Charge-insensitive qubit design derived from the

cooper pair box, Phys. Rev. A **76**, 042319 (2007).

[2] J. Q. You, X. Hu, S. Ashhab, and F. Nori, Low-decoherence flux qubit, Phys. Rev. B **75**, 140515(R) (2007).

[3] U. Vool and M. Devoret, Introduction to quantum electromagnetic circuits, Int. J. Circuit Theory Appl. **45**, 897 (2017).

[4] F. Yan, S. Gustavsson, A. Kamal, J. Birenbaum, A. P. Sears, D. Hover, T. J. Gudmundsen, D. Rosenberg, G. Samach, S. Weber, J. L. Yoder, T. P. Orlando, J. Clarke, A. J. Kerman, and W. D. Oliver, The flux qubit revisited to enhance coherence and reproducibility, Nat. Commun. **7**, 12964 (2016).

[5] M. Steffen, S. Kumar, D. P. DiVincenzo, J. R. Rozen, G. A. Keefe, M. B. Rothwell, and M. B. Ketchen, High-Coherence Hybrid Superconducting Qubit, Phys. Rev. Lett. **105**, 100502 (2010).

[6] J. M. Chow, A. D. Córcoles, J. M. Gambetta, C. Rigetti, B. R. Johnson, J. A. Smolin, J. R. Rozen, G. A. Keefe, M. B. Rothwell, M. B. Ketchen, and M. Steffen, Simple All-Microwave Entangling Gate for Fixed-Frequency Superconducting Qubits, Phys. Rev. Lett. **107**, 080502 (2011).

[7] J. Kelly, R. Barends, A. G. Fowler, A. Megrant, E. Jeffrey, T. C. White, D. Sank, J. Y. Mutus, B. Campbell, Yu Chen, Z. Chen, B. Chiaro, A. Dunsworth, I.-C. Hoi, C. Neill, P. O'Malley, C. Quintana, P. Roushan, A. Vainsencher, J. Wenner, A. N. Cleland, and John M. Martinis, State preservation by repetitive error detection in a superconducting quantum circuit, Nature **519**, 66 (2015).

[8] F. Arute, K. Arya, R. Babbush, D. Bacon, J. C. Bardin, R. Barends, R. Biswas, S. Boixo, F. G. Brandao, D. A. Buell, et al., Quantum supremacy using a programmable superconducting processor, Nature **574**, 505 (2019).

[9] P. S. Mundada, G. Zhang, T. Hazard, and A. A. Houck, Suppression of Qubit Crosstalk in a Tunable Coupling Superconducting Circuit, Phys. Rev. Applied **12**, 054023 (2019).

[10] S. Sheldon, E. Magesan, J. M. Chow, and J. M. Gambetta, Procedure for systematically tuning up cross-talk in the cross-resonance gate, Phys. Rev. A **93**, 060302(R) (2016).

[11] O. Heinsoo, C. K. Andersen, A. Remm, S. Krinner, T. Walter, Y. Salathé, S. Gasparinetti, J.-C. Besse, A. Potočnik, A. Wallraff, and C. Eichler, Rapid high-fidelity multiplexed readout of superconducting qubits, Phys. Rev. Applied **10**, 034040 (2018).

[12] M. Reagor, C. B. Osborn, N. Tezak, A. Staley, et al., Demonstration of universal parametric entangling gates on a multi-qubit lattice, lattice. Sci. Adv. **4**, eaao3603 (2018).

[13] S. A. Caldwell, N. Didier, C. A. Ryan, E. A. Sete, A. Hudson, et al., Parametrically Activated Entangling Gates Using Transmon Qubits, Phys. Rev. Applied **10**, 034050 (2018).

[14] J. Ghosh, A. Galiautdinov, Z. Zhou, A. N. Korotkov, J. M. Martinis, and M. R. Geller, High-fidelity controlled-$\sigma^Z$ gate for resonator-based superconducting quantum computers, Phys. Rev. A **87**, 022309 (2013).

[15] L. H. Pedersen, N. M. Møller, and K. Mølmer, Fidelity of quantum operations, Phys. Lett. A **367**, 47 (2007).

[16] E. Zahedinejad, J. Ghosh, and B. C. Sanders, Designing High-Fidelity Single-Shot Three-Qubit Gates: A Machine-Learning Approach, Phys. Rev. Applied **6**, 054005 (2016).

[17] E. Barnes, C. Arenz, A. Pitchford, and S. E. Economou, Fast microwave-driven three-qubit gates for cavity-coupled superconducting qubits, Phys. Rev. B **96**, 024504 (2017).

[18] R. Barends, C. M. Quintana, A. G. Petukhov, Y. Chen, D. Kafri, et al., Diabatic Gates for Frequency-Tunable Superconducting Qubits, Phys. Rev. Lett. **123**, 210501 (2019).

[19] E. A. Sete, W. J. Zeng, and C. T. Rigetti, A functional architecture for scalable quantum computing, 2016 IEEE International Conference on Rebooting Computing (ICRC).IEEE, 2016: 1-6.

[20] B. Foxen, C. Neill, A. Dunsworth, P. Roushan, B. Chiaro, et al., Demonstrating a Continuous Set of Two-qubit Gates for Near-term Quantum Algorithms, arXiv:2001.08343 (2020).

[21] Y. Chen, C. Neill, P. Roushan, N. Leung, M. Fang, R. Barends, J. Kelly, B. Campbell, Z. Chen, B. Chiaro, A. Dunsworth, E. Jeffrey, A. Megrant, J. Y. Mutus, P. J. J. O'Malley, C. M. Quintana, D. Sank, A. Vainsencher, J. Wenner, T. C. White, M. R. Geller, A. N. Cleland, and J. M. Martinis, Qubit architecture with high coherence and fast tunable coupling, Phys. Rev. Lett. **113**, 220502 (2014).

[22] C. Neill. A path towards quantum supremacy with superconducting qubits. PhD thesis, University of California Santa Barbara, Dec 2017.

[23] F. Yan, P. Krantz, Y. Sung, M. Kjaergaard, D. L. Campbell, T. P. Orlando, S. Gustavsson, and W. D. Oliver, Tunable Coupling Scheme for Implementing High-Fidelity Two-Qubit Gates, Phys. Rev. Applied **10**, 054062 (2018).

[24] L. DiCarlo, J. M. Chow, J. M. Gambetta, L. S. Bishop, B. R. Johnson, D. I. Schuster, J. Majer, A. Blais, L. Frunzio, S. M. Girvin, and R. J. Schoelkopf, Demonstration of two-qubit algorithms with a superconducting quantum processor, Nature(London) **460**, 240 (2009).

[25] R. Krishnan and J. A. Pople, Approximate fourth-order perturbation theory of the electron correlation energy, Int. J. Quantum Chem. **14**, 91 (1978).

[26] R. Winkler, Spin-Orbit Coupling Effects in Two-Dimensional Electron and Hole System (Springer, 2003).




\begin{thebibliography}{99}%

\bibitem{R1} R. Barends, J. Kelly, A. Megrant, A. Veitia, D. Sank, E. Jeffrey, T. C. White, J. Mutus, A. G. Fowler, B. Campbell, Y. Chen, Z. Chen, B. Chiaro, A. Dunsworth, C. Neill, P. O'Malley, P. Roushan, A. Vainsencher, J. Wenner, A. N. Korotkov, A. N. Cleland, and J. M. Martinis, Superconducting quantum circuits at the surface code threshold for fault tolerance, \href{https://www.nature.com/articles/nature13171}{Nature \textbf{508}, 500 (2014)}.

\bibitem{R2} F. Arute, K. Arya, R. Babbush, D. Bacon, J. C. Bardin, R. Barends, R. Biswas, S. Boixo, F. G. Brandao, D. A. Buell, et al., Quantum supremacy using a programmable superconducting processor, \href{https://www.nature.com/articles/s41586%20019%201666%205}{Nature \textbf{574}, 505 (2019)}.

\bibitem{R3} M. Kjaergaard, M. E. Schwartz, J. Braum\"{u}ller, P. Krantz, J. I.-J. Wang, S. Gustavsson, and W. D. Oliver, Superconducting qubits: Current state of play, \href{https://arxiv.org/abs/1905.13641}{arXiv:1905.13641 (2019)}.

\bibitem{R4} F. W. Strauch, P. R. Johnson, A. J. Dragt, C. J. Lobb, J. R. Anderson, and F. C. Wellstood, Quantum Logic Gates for Coupled Superconducting Phase Qubits, \href{https://journals.aps.org/prl/abstract/10.1103/PhysRevLett.91.167005}{Phys. Rev. Lett. \textbf{91}, 167005 (2003)}.

\bibitem{R5} L. DiCarlo, J. M. Chow, J. M. Gambetta, L. S. Bishop, B. R. Johnson, D. I. Schuster, J. Majer, A. Blais, L. Frunzio, S. M. Girvin, and R. J. Schoelkopf, Demonstration of two-qubit algorithms with a superconducting quantum processor, \href{https://www.nature.com/articles/nature08121}{Nature (London) \textbf{460}, 240 (2009)}.

\bibitem{R6} J. Koch, T. M. Yu, J. Gambetta, A. A. Houck, D. I. Schuster, J. Majer, A. Blais, M. H. Devoret, S.
    M. Girvin, and R. J. Schoelkopf, Charge-insensitive qubit design derived from the cooper pair box, \href{https://journals.aps.org/prapplied/abstract/10.1103/PhysRevApplied.3.044009}{Phys. Rev. A \textbf{76}, 042319 (2007)}.

\bibitem{R7} M. Steffen, S. Kumar, D. P. DiVincenzo, J. R. Rozen, G. A. Keefe, M. B. Rothwell, and M. B. Ketchen, High-Coherence Hybrid Superconducting Qubit, \href{https://journals.aps.org/prl/abstract/10.1103/PhysRevLett.105.100502}{Phys. Rev. Lett. \textbf{105}, 100502 (2010)}.

\bibitem{R8} J. M. Chow, A. D. C\'{o}rcoles, J. M. Gambetta, C. Rigetti, B. R. Johnson, J. A. Smolin, J. R. Rozen, G. A. Keefe, M. B. Rothwell, M. B. Ketchen, and M. Steffen, Simple All-Microwave Entangling Gate for Fixed-Frequency Superconducting Qubits, \href{https://journals.aps.org/prl/abstract/10.1103/PhysRevLett.107.080502}{Phys. Rev. Lett. \textbf{107}, 080502 (2011)}.

\bibitem{R9} F. Yan, S. Gustavsson, A. Kamal, J. Birenbaum, A. P. Sears, D. Hover, T. J. Gudmundsen, D. Rosenberg, G. Samach, S. Weber, J. L. Yoder, T. P. Orlando, J. Clarke, A. J. Kerman, and W. D. Oliver, The flux qubit revisited to enhance coherence and reproducibility, \href{https://www.nature.com/articles/ncomms12964}{Nat. Commun. \textbf{7}, 12964 (2016)}.

\bibitem{R10} M. A. Rol, F. Battistel, F. K. Malinowski, C. C. Bultink, B. M. Tarasinski, R.Vollmer, N. Haider, N. Muthusubramanian,
    A. Bruno, B. M. Terhal et al., A Fast, Low-Leakage,High-Fidelity Two-Qubit Gate for a Programmable Superconducting
    Quantum Computer, \href{https://journals.aps.org/prl/abstract/10.1103/PhysRevLett.123.210501}{Phys. Rev. Lett. \textbf{123}, 120502 (2019)}

\bibitem{R11} R. Barends, C. M. Quintana, A. G. Petukhov, Y. Chen, D. Kafri, et al., Diabatic Gates for Frequency-Tunable Superconducting Qubits, \href{https://journals.aps.org/prl/abstract/10.1103/PhysRevLett.123.210501}{Phys. Rev. Lett. \textbf{123}, 210501 (2019)}.

\bibitem{R12} B. Foxen, C. Neill, A. Dunsworth, P. Roushan, B. Chiaro, et al., Demonstrating a Continuous Set of Two-qubit Gates for Near-term Quantum Algorithms, \href{https://arxiv.org/abs/2001.08343}{arXiv:2001.08343 (2020)}.

\bibitem{R13} S. Sheldon, E. Magesan, J. M. Chow, and J. M. Gambetta, Procedure for systematically tuning up cross-talk in the cross-resonance gate, \href{https://journals.aps.org/pra/abstract/10.1103/PhysRevA.93.060302}{Phys. Rev. A \textbf{93}, 060302(R) (2016)}.

\bibitem{R14} A.D. Patterson, J. Rahamim, T. Tsunoda, P.A. Spring, S. Jebari, K. Ratter, M. Mergenthaler, G. Tancredi, B. Vlastakis, M. Esposito, and P.J. Leek, Calibration of a Cross-Resonance Two-Qubit Gate Between Directly Coupled Transmons, \href{https://journals.aps.org/prapplied/abstract/10.1103/PhysRevApplied.12.064013}{Phys. Rev. Applied \textbf{12}, 064013 (2019)}

\bibitem{R15} D. C. McKay, S. Filipp, A. Mezzacapo, E. Magesan, J. M. Chow, and J. M. Gambetta, Universal gate for fixed-frequency qubits via a tunable bus, \href{https://journals.aps.org/prapplied/abstract/10.1103/PhysRevApplied.6.064007}{Phys. Rev. Applied \textbf{6}, 064007 (2016)}.

\bibitem{R16} M. Reagor, C. B. Osborn, N. Tezak, A. Staley, et al., Demonstration of universal parametric entangling gates on a multi-qubit lattice, \href{https://advances.sciencemag.org/content/4/2/eaao3603?utm_source=TrendMD&utm_medium=cpc&utm_campaign=TrendMD_1}{lattice. Sci. Adv. \textbf{4}, eaao3603 (2018)}.

\bibitem{R17} S. A. Caldwell, N. Didier, C. A. Ryan, E. A. Sete, A. Hudson, et al., Parametrically Activated Entangling Gates Using Transmon Qubits, \href{https://journals.aps.org/prapplied/abstract/10.1103/PhysRevApplied.10.034050}{Phys. Rev. Applied \textbf{10}, 034050 (2018)}.


\bibitem{R18} J. M. Gambetta, A. D. C\'{o}rcoles, S. T. Merkel, B. R. Johnson,
    J. A. Smolin, J. M. Chow, C. A. Ryan, C. Rigetti, S. Poletto, T. A. Ohki, M. B. Ketchen, and M. Steffen, Characterization of Addressability by Simultaneous Randomized Benchmarking, \href{https://journals.aps.org/prapplied/abstract/10.1103/PhysRevApplied.10.034050}{Phys. Rev. Lett. \textbf{109}, 240504 (2012)}.

\bibitem{R19} D. C. McKay, S. Sheldon, J. A. Smolin, J. M. Chow, and J. M. Gambetta, Three Qubit Randomized Benchmarking, \href{https://journals.aps.org/prl/abstract/10.1103/PhysRevLett.122.200502}{Phys. Rev. Lett. \textbf{122}, 200502 (2019)}.

\bibitem{R20} M. Takita, A. W. Cross, A. D. C\'{o}rcoles, J. M. Chow, and J. M. Gambetta, Experimental Demonstration
    of Fault-tolerant State Preparation with Superconducting Qubits, \href{https://journals.aps.org/prl/abstract/10.1103/PhysRevLett.119.180501}{ Phys. Rev. Lett. \textbf{119}, 180501 (2017)}.


\bibitem{R21} M. Takita, A. D. C\'{o}rcoles, E. Magesan, B. Abdo, M. Brink, A. Cross, J. M. Chow, and J. M. Gambetta, Demonstration of Weight-Four Parity Measurements in the Surface Code Architecture, \href{https://journals.aps.org/prl/abstract/10.1103/PhysRevLett.117.210505}{Phys. Rev. Lett. \textbf{117}, 210505 (2016)}.

\bibitem{R22} C. K. Andersen, A. Remm, S. Balasiu, S. Krinner, J. Heinsoo, J. Besse, M. Gabureac, A. Wallraff, and C. Eichler, Entanglement Stabilization using Parity Detection and Real-Time Feedback in Superconducting Circuits, \href{https://arxiv.org/abs/1902.06946}{arXiv:1902.06946 (2019)}.

\bibitem{R23} C. C. Bultink, T. E. O'Brien, R. Vollmer, N. Muthusubramanian, M. W. Beekman, M. A. Rol, X. Fu, B. Tarasinski, V. Ostroukh, B. Varbanov, A. Bruno, L. DiCarlo, Protecting quantum entanglement from qubit errors and leakage via repetitive parity measurements, \href{https://arxiv.org/abs/1905.12731}{arXiv:1905.12731 (2020)}.

\bibitem{R24} C. K. Andersen, A. Remm, S. Lazar, S. Krinner, N. Lacroix, G. J. Norris, M. Gabureac, C. Eichler, and A. Wallraff, Repeated Quantum Error Detection in a Surface Code, \href{https://arxiv.org/abs/1912.09410}{arXiv:1912.09410 (2020)}.

\bibitem{R25} A. G. Fowler, M. Mariantoni, J. M. Martinis, and A. N. Cleland, Surface codes: Towards practical large-scale quantum computation,
    \href{https://journals.aps.org/pra/abstract/10.1103/PhysRevA.86.032324}{Phys. Rev. A \textbf{86}, 032324 (2012)}.

\bibitem{R26} See Supplemental Material: Circuit Hamiltonian, which includes Refs.\cite{R27,R28,R29,R30}.

\bibitem{R27} J. Q. You, X. Hu, S. Ashhab, and F. Nori, Low-decoherence flux qubit, \href{https://journals.aps.org/prb/abstract/10.1103/PhysRevB.75.140515}{Phys. Rev. B \textbf{75}, 140515(R) (2007)}.

\bibitem{R28} U. Vool and M. Devoret, Introduction to quantum electromagnetic circuits, \href{https://onlinelibrary.wiley.com/doi/abs/10.1002/cta.2359}{Int. J. Circuit Theory Appl. \textbf{45}, 897 (2017)}.

\bibitem{R29} J. Kelly, R. Barends, A. G. Fowler, A. Megrant, E. Jeffrey, T. C. White, D. Sank, J. Y. Mutus, B. Campbell, Yu Chen, Z. Chen, B. Chiaro, A. Dunsworth, I.-C. Hoi, C. Neill, P. O'Malley, C. Quintana, P. Roushan, A. Vainsencher, J. Wenner, A. N. Cleland, and John M. Martinis, State preservation by repetitive error detection in a superconducting quantum circuit, \href{https://www.nature.com/articles/nature14270}{Nature \textbf{519}, 66 (2015)}.

\bibitem{R30} O. Heinsoo, C. K. Andersen, A. Remm, S. Krinner, T. Walter, Y. Salath\'{e}, S. Gasparinetti, J.-C. Besse, A. Poto\v{c}nik, A. Wallraff, and C. Eichler, Rapid high-fidelity multiplexed readout of superconducting qubits, \href{https://journals.aps.org/prapplied/abstract/10.1103/PhysRevApplied.10.034040}{Phys. Rev. Applied \textbf{10}, 034040 (2018)}.

\bibitem{R31} See Supplemental Material: Triple degeneracy point.

\bibitem{R32} See Supplemental Material: Qubit coupled via a coupler, which includes Refs.\cite{R33,R34}.

\bibitem{R33} R. Krishnan and J. A. Pople, Approximate fourth-order perturbation theory of the electron correlation
    energy, \href{https://onlinelibrary.wiley.com/doi/abs/10.1002/qua.560140109}{Int. J. Quantum Chem. \textbf{14}, 91 (1978)}.

\bibitem{R34} R. Winkler, Spin-Orbit Coupling Effects in Two-Dimensional Electron and Hole System (Springer, 2003).

\bibitem{R35} See Supplemental Material: Gate operation with diabatic scheme, which includes Refs.\cite{R36,R37}.

\bibitem{R36}  E. Zahedinejad, J. Ghosh, and B. C. Sanders, Designing High-Fidelity Single-Shot Three-Qubit Gates: A Machine-Learning Approach, \href{https://journals.aps.org/prapplied/abstract/10.1103/PhysRevApplied.6.054005}{Phys. Rev. Applied \textbf{6}, 054005  (2016)}.

\bibitem{R37}  E. Barnes, C. Arenz, A. Pitchford, and S. E. Economou, Fast microwave-driven three-qubit gates for cavity-coupled superconducting qubits, \href{https://journals.aps.org/prb/abstract/10.1103/PhysRevB.96.024504}{Phys. Rev. B \textbf{96}, 024504 (2017)}.

\bibitem{R38} J. Ghosh, A. Galiautdinov, Z. Zhou, A. N. Korotkov, J. M. Martinis, and M. R. Geller, High-fidelity controlled-$\sigma^{Z}$ gate for resonator-based superconducting quantum computers,
    \href{https://journals.aps.org/pra/abstract/10.1103/PhysRevA.87.022309}{Phys. Rev. A \textbf{87}, 022309 (2013)}.

\bibitem{R39} L. H. Pedersen, N. M. M{\o}ller, and K. M{\o}lmer, Fidelity of quantum operations,
    \href{https://journals.aps.org/pra/abstract/10.1103/PhysRevA.87.022309}{Phys. Lett. A \textbf{367}, 47 (2007)}.

\bibitem{R40} We note that similar to the CZ gate case, it is nearly impossible to simultaneously minimize the swap error and leakage for iSWAP gates in present system with fixed coupling strength, as shown in Fig.$\,$4(e). In present work, hold time is chosen to
    minimize the swap error for the implementation of the iSWAP gate.

\bibitem{R41} Two theoretical superconducting architectures with multi-type qubits were previously proposed, but they address different challenges \cite{R42,R43}. In the work of M. Elliott \emph{et al.} \cite{R42}, two qubits with opposite anharmonicities are coupled to a cavity to achieve cancelation of the cavity self-Kerr effect. In the work of E. A. Sete \emph{et al.} \cite{R43}, transmon qubits are coupled to fluxoinum in a lattice with an -A-B-A-B- pattern, where ZZ coupling resulted from the interaction between $|11\rangle$ and $|02\rangle$ (Here the first digit denotes the transmon states and the second one denotes the fluxonium states) is suppressed by the strong fluxonium anharmonicity, while the contribution of the inetraction between $|11\rangle$ and $|20\rangle$ is preserved.

\bibitem{R42} M. Elliott, J. Joo, and E. Ginossar, Designing Kerr interactions using multiple superconducting qubit types in a single circuit, \href{https://iopscience.iop.org/article/10.1088/1367-2630/aa9243/meta}{New. J. Phys. \textbf{20}, 023037 (2018)}.

\bibitem{R43} E. A. Sete, W. J. Zeng, and C. T. Rigetti, A functional architecture for scalable quantum computing, 2016 IEEE International Conference on Rebooting Computing (ICRC).\href{https://ieeexplore.ieee.org/abstract/document/7738703}{IEEE, 2016: 1-6}.

\bibitem{R44} D. M. Abrams, N. Didier, B. R. Johnson, M. P. da Silva, C. A. Ryan, Implementation of the XY interaction family with calibration of a single pulse, \href{https://arxiv.org/abs/1912.04424}{arXiv:1912.04424 (2019)}.

\bibitem{R45} Y. Chen, C. Neill, P. Roushan, N. Leung, M. Fang, R. Barends, J. Kelly, B. Campbell, Z. Chen, B. Chiaro, A. Dunsworth, E. Jeffrey, A. Megrant, J. Y. Mutus, P. J. J. O'Malley, C. M. Quintana, D. Sank, A. Vainsencher, J. Wenner, T. C. White, M. R. Geller, A. N. Cleland, and J. M. Martinis, Qubit architecture with high coherence and fast tunable coupling,     \href{https://journals.aps.org/pra/abstract/10.1103/PhysRevA.87.022309}{Phys. Rev. Lett. \textbf{113}, 220502 (2014)}.

\bibitem{R46} C. Neill. A path towards quantum supremacy with superconducting qubits. PhD thesis, University of California Santa Barbara, Dec 2017.

\bibitem{R47} F. Yan, P. Krantz, Y. Sung, M. Kjaergaard, D. L. Campbell, T. P. Orlando, S. Gustavsson, and W. D. Oliver, Tunable Coupling Scheme for Implementing High-Fidelity Two-Qubit Gates, \href{https://journals.aps.org/prapplied/abstract/10.1103/PhysRevApplied.10.054062}{Phys. Rev. Applied \textbf{10}, 054062 (2018)}.

\bibitem{R48} P. S. Mundada, G. Zhang, T. Hazard, and A. A. Houck, Suppression of Qubit Crosstalk in a Tunable Coupling Superconducting Circuit, \href{http://dx.doi.org/10.1103/PhysRevApplied.12.054023}{Phys. Rev. Applied \textbf{12}, 054023 (2019)}.



\end{thebibliography}
\end{document}